\pdfoutput=1

\documentclass[twocolumn]{aastex631}

\received{December 22, 2023}
\revised{March 4, 2024}
\accepted{\today}

\submitjournal{AJ}

\usepackage{graphicx}	
\usepackage{amsmath}	
\usepackage{amssymb}	
\usepackage{booktabs}	

\shorttitle{Modeling Noble Gas Emission Lines in T Cha's Disk Wind}
\shortauthors{Sellek et al.}

\begin{document}

\title{Modeling JWST MIRI-MRS Observations of T Cha: Mid-IR Noble Gas Emission Tracing a Dense Disk Wind}

\correspondingauthor{Andrew Sellek}
\email{sellek@strw.leidenuniv.nl}

\author[0000-0003-0330-1506]{Andrew D. Sellek}
\affiliation{Institute of Astronomy, University of Cambridge, Madingley Road, Cambridge CB3 0HA, UK}
\affiliation{Leiden Observatory, Leiden University, 2300 RA Leiden, The Netherlands}

\author[0000-0003-3401-1704]{Naman S. Bajaj}
\affiliation{Lunar and Planetary Laboratory, The University of Arizona, Tucson, AZ 85721, USA}

\author[0000-0001-7962-1683]{Ilaria Pascucci}
\affiliation{Lunar and Planetary Laboratory, The University of Arizona, Tucson, AZ 85721, USA}

\author{Cathie J. Clarke}
\affiliation{Institute of Astronomy, University of Cambridge, Madingley Road, Cambridge CB3 0HA, UK}

\author[0000-0001-6410-2899]{Richard Alexander}
\affiliation{School of Physics and Astronomy, University of Leicester, University Road, Leicester, LE1 7RH, UK}

\author[0000-0001-8184-5547]{Chengyan Xie}
\affiliation{Lunar and Planetary Laboratory, The University of Arizona, Tucson, AZ 85721, USA}

\author[0000-0002-4687-2133]{Giulia Ballabio}
\affiliation{Astrophysics Group, Department of Physics, Imperial College London, Prince Consort Rd, London SW7 2AZ, UK }

\author[0000-0003-0777-7392]{Dingshan Deng}
\affiliation{Lunar and Planetary Laboratory, The University of Arizona, Tucson, AZ 85721, USA}

\author{Uma Gorti}
\affiliation{NASA Ames Research Center, Moffett field, CA 94035, USA}
\affiliation{Carl Sagan Center, SETI Institute, Mountain View, CA 94943, USA}

\author[0000-0001-8612-3236]{Andras Gaspar}
\affiliation{Steward Observatory, The University of Arizona, Tucson, AZ 85721, USA}

\author{Jane Morrison}
\affiliation{Steward Observatory, The University of Arizona, Tucson, AZ 85721, USA}



\begin{abstract}
\,[\ion{Ne}{2}] $12.81\,\mu\mathrm{m}$ emission is a well-used tracer of protoplanetary disk winds due to its blueshifted line profile.
MIRI-MRS recently observed T Cha, detecting this line along with lines of [\ion{Ne}{3}], [\ion{Ar}{2}] and [\ion{Ar}{3}], with the [\ion{Ne}{2}] and [\ion{Ne}{3}] lines found to be extended while the [\ion{Ar}{2}] was not.
In this complementary work, we use these lines to address long-debated questions about protoplanetary disk winds regarding their mass-loss rate, the origin of their ionization, and the role of magnetically-driven winds as opposed to photoevaporation.
To this end, we perform photoionization radiative transfer on simple hydrodynamic wind models to map the line emission. We compare the integrated model luminosities to those observed with MIRI-MRS to identify which models most closely reproduce the data and produce synthetic images from these to understand what information is captured by measurements of the line extents.
Along with the low degree of ionization implied by the line ratios, the relative compactness of [\ion{Ar}{2}] compared to [\ion{Ne}{2}] is particularly constraining. This requires \ion{Ne}{2} production by hard X-rays and \ion{Ar}{2} production by soft X-rays (and/or EUV) in an extended ($\gtrsim10\,\mathrm{au}$) wind that is shielded from soft X-rays - necessitating a dense wind with material launched on scales down to $\sim1\,\mathrm{au}$.
Such conditions could be produced by photoevaporation, whereas an extended MHD wind producing equal shielding would likely underpredict the line fluxes. However, a tenuous inner MHD wind may still contribute to shielding the extended wind.
This picture is consistent with constraints from spectrally-resolved line profiles.
\end{abstract}



\section{Introduction} \label{sec:intro}

Protoplanetary disks are found around stars in the first few Myr of their lifetimes \citep[with estimates of the average disk lifetime ranging from $2-8\,\mathrm{Myr}$,][]{Pfalzner_2022}. Despite the importance of this timescale for understanding planet formation, there remain many open questions about the processes driving the evolution and ultimate dispersal of these disks, including the role played by different wind mechanisms during different phases of the disk's evolution.
Disk winds have been detected from disks across all stages of evolution \citep[see][for a recent review]{Pascucci_2023} using emission from atomic and/or molecular species, with mass-loss rates estimated to be comparable or even greater than accretion rates onto the central star.
Magnetically-driven winds have also been attracting increased attention in recent years as a mechanism to drive accretion \citep[see][for a recent review on how this impacts on disk demographics]{Manara_2023}.

Winds may be launched from protoplanetary disks due to heating by high-energy radiation from the central star or by the action of magnetic fields. In the former case - known as photoevaporation - winds are only expected to originate from radii $\gtrsim 0.1\,r_G$ \citep{Liffman_2003,Font_2004,Dullemond_2007,Alexander_2014,Clarke_2016} where $r_G = \frac{GM_*}{c_{\rm S}^2}$ is the ``gravitational radius'' where the thermal energy becomes great enough to overcome the gravitational potential of the star \citep{Hollenbach_1994}. The exact location of this radius depends on the temperature reached by the heated gas; consequently, Far Ultraviolet (FUV) radiation produces mass-loss concentrated more towards the outer disk \citep{Gorti_2015} than X-ray or Extreme Ultraviolet (EUV) radiation.
Conversely, \textit{cold} magnetohydrodynamic (MHD) winds, launched by either the stellar \citep{Blandford_1982,Pelletier_1992,Wardle_1993,Ferreira_1997} or disk \citep{Suzuki_2009} magnetic field, are powered by the gravitational energy lost by accreting material \citep[e.g.][]{Ferreira_1995,Suzuki_2016} and/or the magnetic field energy \citep{Lesur_2021} and so may originate at any radius.
An intermediate class of \textit{magnetothermal} winds \citep{Bai_2017} where both thermal energy and magnetic forces contribute to driving the wind is also possible \citep{Wang_2019,Gressel_2020,Rodenkirch_2020,Sarafidou_2023}.
Thus, constraining the launch radius of winds observationally can help distinguish between these scenarios. 

Sensitive Atacama Large Millimetre/Submillimetre Array (ALMA) observations have spatially resolved wide-angle outflows consistent with a wind at all stages of protostellar evolution. These include Class 0 sources HH 212 \citep{Lee_2021} and L1448-mm \citep{Nazari_2024}, DG Tau B \citep[Class I;][]{deValon_2020}, HL Tau \citep[Class I-II;][]{Klaassen_2016} and Class II sources DG Tau A \citep{AgraAmboage_2014,Gudel_2018} \& HH 30 \citep{Louvet_2018}.
However for the majority of systems, we lack the sensitivity to robustly establish a wind origin for large scale emission and trace it back to the disk.

On the other hand, spectrally resolved atomic emission lines frequently show blueshifts which indicate material flowing away from the disk \citep[e.g.][]{Hartigan_1995}.
High-resolution spectroscopy allows line profiles to be decomposed into several components which are typically classified based on their centroid as high-velocity components (HVC $>30\,\mathrm{km\,s^{-1}}$) - thought to be associated with jets - or low-velocity components (LVC $<30\,\mathrm{km\,s^{-1}}$) - thought to trace winds.
The best-studied line is the [\ion{O}{1}] 6300 \r{A}. Its LVC is often subdivided into two components \citep[e.g.][]{Rigliaco_2013,Simon_2016} - one in order to give a narrow central peak (the Narrow Component, NC/NLVC) and a second to capture the line wings (the Broad Component, BC/BLVC).
Assuming Keplerian line broadening, the BLVC maps onto a launch radius of $0.05-0.5\,\mathrm{au}$ while that of the NLVC lies mostly at $\gtrsim \mathbf{0.5}\,\mathrm{au}$ \citep{Simon_2016,McGinnis_2018}. This is typically taken as evidence for the origin of the BLVC in a magnetically-driven inner disk wind \citep{Simon_2016}.
The kinematics of the NLVC correlate closely with those of the BLVC \citep{Banzatti_2019} which may mean that it traces a more-extended part of the same wind although \citet{Weber_2020} suggest that correlations with a third variable (such as accretion luminosity) may also explain this.

A more promising emission line for tracing an extended, possibly photoevaporative, wind is the [\ion{Ne}{2}] 12.81 $\mu$m line \citep[e.g.][]{Pascucci_2020}. This also shows a blueshift \citep{Herczeg_2007,Pascucci_2009} and usually shows either an HVC or a single narrow LVC \citep{Sacco_2012,Baldovin-Saavedra_2012,Pascucci_2020}, the latter suggesting an origin further out in the disk than [\ion{O}{1}].
This scneario is also supported by the fact that it remains blueshifted even in disks with large cavities where the [\ion{O}{1}] emission may have much smaller or no blueshift \citep[e.g. TW Hya -][]{Pascucci_2011,Fang_2023b}, and therefore likely has to originate further out, outside the cavity.
Further evidence that the [\ion{Ne}{2}] traces wind material distinct to that traced by the [\ion{O}{1}] is that former gets brighter as the spectral index between 13 and 31 $\mu\mathrm{m}$ increases (which indicates clearing of dust from the inner disk), while latter gets fainter \citep{Pascucci_2020}.

In the context of photoevaporation, most previous modeling of these observational diagnostics relied on dedicated hydrodynamical models which were then each post-processed \citep{Font_2004,Alexander_2008,Ercolano_2010,Ercolano_2016,Picogna_2019,Weber_2020,Rab_2022,Rab_2023}. The development of self-similar solutions for thermal disk winds by \citet{Clarke_2016} and their subsequent extension to the more realistic case of elevated wind bases \citep{Sellek_2021} has enabled the much easier interpretation of observations. The density and velocity fields may be generated easily and then scaled to a chosen normalization (typically of the sound speed and integrated mass-loss rate). This allows larger parameter spaces to be explored and individual sources to be modeled without requiring dedicated hydrodynamical models for each.

Self-similar models were first applied in such a way by \citet{Ballabio_2020} to model a sample of [\ion{O}{1}] 6300 \r{A} \citep{Banzatti_2019} and [\ion{Ne}{2}] $12.81~\mu\mathrm{m}$ line profiles.
In order to convert the densities provided by self-similar models to line profiles they assumed spatially constant elemental abundances, ionization fractions and temperatures. They then normalize their profiles to avoid the uncertainty introduced by the unknown magnitudes of these quantities.
Their results suggest that the two key line profile parameters - the centroid shift and the FWHM - are highly sensitive to the wind's sound speed.
Comparing to the observed data, they found that the [\ion{Ne}{2}] centroids and FWHM both preferred a fast wind with $c_{\rm S}\sim10\,\mathrm{km~s^{-1}}$ as appropriate to an EUV wind, while the smaller [\ion{O}{1}] blueshifts were more consistent with a slower wind with $c_{\rm S}\sim3-5\,\mathrm{km~s^{-1}}$.

While the work of \citet{Ballabio_2020} supports the hypothesis of different origins for the [\ion{Ne}{2}] and [\ion{O}{1}] emission \citep{Pascucci_2020}, the assumptions of uniform temperatures and ionic abundances prevent a single model from capturing these differences. For example, by postprocessing hydrodynamical simulation of an X-ray--driven photoevaporative wind using a photoionization code, \citet{Ercolano_2016} showed that the inner regions may be EUV-heated and therefore hotter and more ionized. This leads to emission from different lines probing different regions of the wind. Thus, understanding the location and extent of the emitting regions is crucial to determining wind launch radii from observations.

In this work we therefore combine the strengths of these approaches - the ease of use of the self-similar models with the realistic thermochemical structure produced by photoionization codes - to study emission lines from photoevaporative winds. In line with previous works, we model line fluxes and shapes. Moreover, since it is becoming possible for the first time to spatially resolve these lines, we produce the first synthetic images of JWST observations of our lines of interest.
To achieve this, we post-process self-similar wind models (for a variety of mass-loss rates and wind extents) using the Monte Carlo photoionization code \textsc{mocassin} and produce more realistic models of the emitting regions of different lines; our full methodology is set out in Section \ref{sec:methods}.
We focus on the [\ion{Ne}{2}] 12.81 $\mu$m line, along with the [\ion{Ne}{3}] 15.55 $\mu$m, [\ion{Ar}{2}] 6.98 $\mu$m and [\ion{Ar}{3}] 8.99 $\mu$m lines. This combination allows for the ionization state of the emitting gas and the shape of the ionizing spectrum to be constrained \citep{Hollenbach_2009b}.

Importantly, these lines of Ne and Ar can be observed with the Medium Resolution Spectrometer (MRS) Integral Field Units (IFUs) onboard the Mid-Infrared Instrument (MIRI) of JWST.
This instrument has observed the nearby \citep[102.7 pc,][]{Gaia_2022} protoplanetary disk around T Cha \citep[under Program GO 2260,][]{Pascucci_2021}, resulting in the first simultaneous detection of all four lines in a protoplanetary disk, along with the determination of extended emission in the [\ion{Ne}{2}] 12.81 $\mu$m line and [\ion{Ne}{3}] 15.55 $\mu$m lines \citep{Bajaj_2024}.

We thus focus our modeling effort on interpreting these observations of T Cha and testing to what extent photoevaporation can explain the data.
In Section \ref{sec:ratios} we compare predicted line luminosities and ratios to the observations.
In Section \ref{sec:images} we present synthetic images made from our most promising models using the MIRISim package \citep{Klaassen_2021} and analyze these in the same way as the observations in order to understand what they can tell us about the wind's structure.
In Section \ref{sec:discussion} we discuss our results in comparison to further data on the system - including line profiles at high spectral resolution - and place them in the wider context of constraints on wind launching.
Finally we summarize our conclusions in Section \ref{sec:conclusions}.

\section{Model description} \label{sec:style}
\label{sec:methods}

\begin{table*}[t]
        \centering
        \caption{Properties of the T Cha System}
        \label{tab:TCha}
        \begin{tabular}{rlll}
        \tableline
        Property    & Value \& Unit          & Comment           & Reference \\
        \tableline
        $M_*$       & 1.5 $M_{\odot}$       &                   & \citet{Olofsson_2011} \\
        Distance    & $102.7 \pm 0.3$ pc    & Gaia DR3 Value    & \citet{Gaia_2022} \\
        Inclination & $73^\circ$            & Measured using $3$ mm continuum  & \citet{Hendler_2018}  \\
        $R_{\rm cav}$ (mm)   & 33 au     &                   & \citet{Francis_2020} \\
        $R_{\rm cav}$ ($\mu{\rm m}$)   & 15 au     &                   & \citet{Xie_2023} \\
        $R_{\rm out}$ (mm)      & 44 au     &                   & \citet{Francis_2020} \\
        $R_{\rm out}$ ($\mu{\rm m}$)      & 58 au     &                   & \citet{Pohl_2017} \\
        $R_{\rm out}$ (gas)     & 220 au    & Measured in \textsuperscript{12}CO              & \citet{Huelamo_2015} \\
        $\dot{M}_{\rm acc}$     & $10^{-8.4}\,M_{\odot}\,\mathrm{yr}^{-1}$ & Using [\ion{O}{1}]-$L_{\rm acc}$ relationship \citep{Nisini_2018} & \citet{Cahill_2019} \\
        \tableline
        \end{tabular}
        \tablecomments{Where appropriate, values have been rescaled from the original references to use the Gaia DR3 distance. $R_{\rm cav}$ refers to the outer radius of the dust cavity as measured at each wavelength regime.}
\end{table*}

\subsection{Constructing the density grids}
Given a mass of $M_*=1.5\,M_\odot$ and assuming a maximal wind temperature of $T=10^4\,\mathrm{K}$ (equivalent to a sound speed of $c_{\rm S}\approx10\,\mathrm{km\,s^{-1}}$), the gravitational radius for T Cha is $r_G := \frac{GM_*}{c_S^2} \approx 13\,\mathrm{au}$. This is the typical lengthscale outside of which the gravitational potential well of the star does not significantly impede the launch of a wind.

We use the self-similar solutions of \citet{Sellek_2021} to produce models of the wind's density and velocity structure.
These provide normalized densities $\tilde{\rho}$ and velocities $\tilde{u}$ along each streamline, defined according to
\begin{align}
    \rho &= \rho_{\rm b} \tilde{\rho} \\
    u    &= \mathcal{M}_{\rm b} c_{\rm S, b} \tilde{u}
\end{align}
where $\rho_{\rm b}$, $\mathcal{M}_{\rm b}$ and $c_{\rm S, b}$ are the density, Mach number and sound speed at the base of the streamline.

\begin{figure*}[p]
    \centering
    \includegraphics[angle=90,origin=c,height=\linewidth]{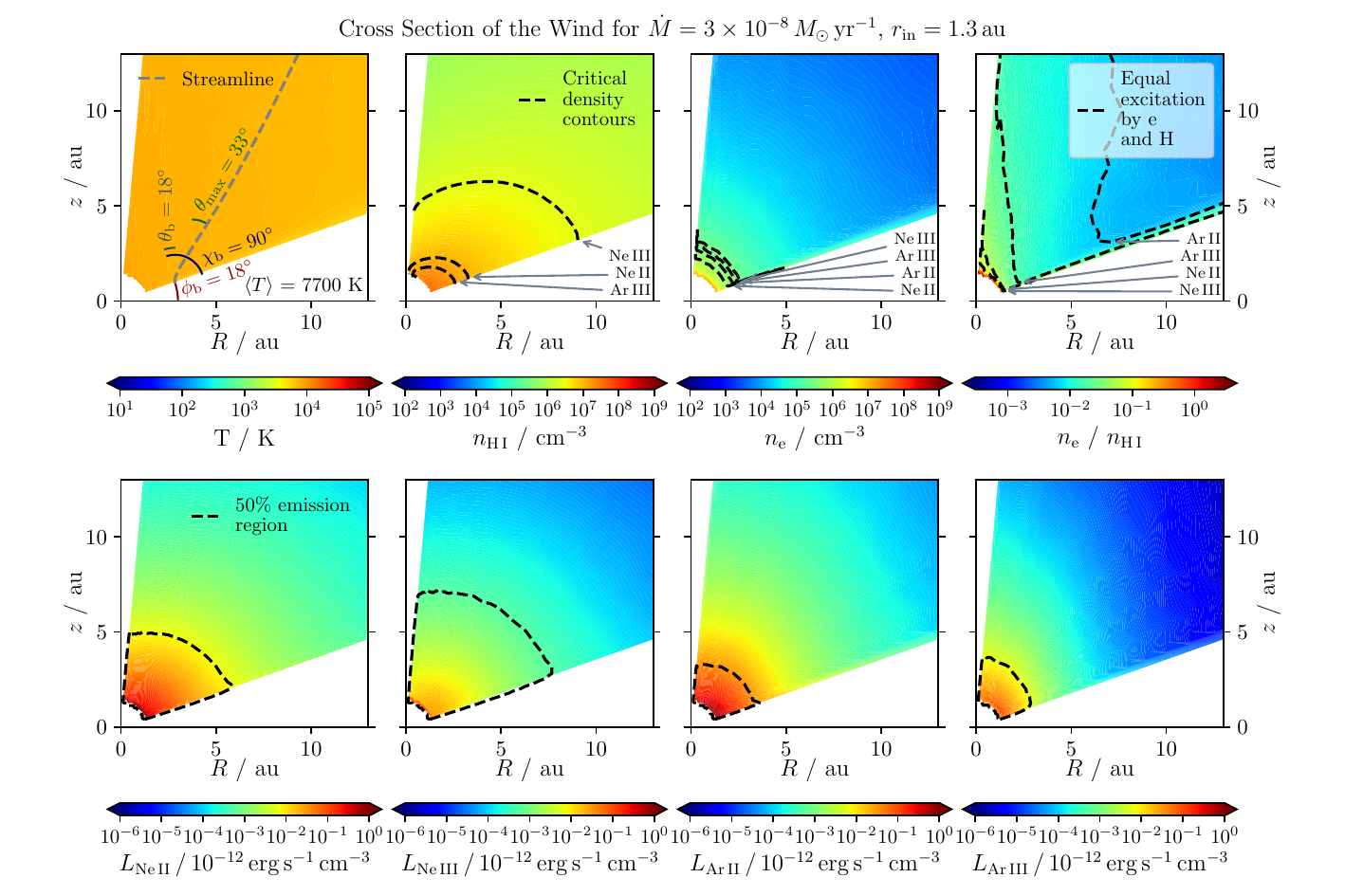}
    \caption{Cross sections of the inner 13 au ($1\,r_G$) of the wind. The first panel shows a streamline from the self-similar model (gray dashed line) superposed on the temperatures calculated by \textsc{mocassin}, with various significant angles indicated. The second and third panels shows the neutral hydrogen density (at this level of ionization, essentially equal to the gas density) and electron densities respectively; the dashed contours show where the critical densities with neutral H/electrons are reached for each transition and are labelled by the corresponding ion. The fourth panel shows ionization fraction with the contours indicating where neutral H and electrons should conribute equally to excitation of the transitions. The bottom row shows, from left to right, the emissivities of the [\ion{Ne}{2}], [\ion{Ne}{3}], [\ion{Ar}{2}] and [\ion{Ar}{3}] lines; in each case the contour encloses the brightest region where 50 per cent of the total luminosity is emitted.}
    \label{fig:Rz_maps}
\end{figure*}

These quantities depend on four key parameters:
\begin{itemize}
    \item \textbf{The slope of the density at the base of the wind, $\rho_{\rm b} \propto r^{-b}$}. In such a model, the column density is accumulated at small radii for $b>1$ and large radii for $b\leq1$. We focus on $b=1.5$, a value motivated both by theory/simulations \citep{Hollenbach_1994,Picogna_2019} as well as previous comparisons to observations of TW Hya \citep{Pascucci_2011,Ballabio_2020}. Note that this is also the scaling adopted in various self-similar magnetized wind models \citep{Blandford_1982,Lesur_2021}. In Section \ref{sec:profiles} we justify this further by showing a lower $b$ would produce line profiles that are too narrow.
    \item \textbf{The slope of the temperature profile in the wind, $T \propto r^{-\tau}$}. We assume this to be $\tau=0$, i.e. an isothermal case. This is typical of EUV-heated winds, but nearly-isothermal conditions are also seen in X-ray--heated winds \citep{Picogna_2019,Picogna_2021} and \citet{Nakatani_2018a} also found $\tau=0.319$ at solar metallicity for FUV-heated winds. Such small temperature gradients only weakly affect the density and velocity structure \citep{Sellek_2021}, changing $\mathcal{M}_{\rm b}$ by no more than $\sim10$ per cent. The results of our \textsc{mocassin} calculations also show {near isothermal temperatures throughout most of their volume (Figure \ref{fig:Rz_maps}); therefore neglecting $\tau \neq 0$ is a small source of uncertainty.
    \item \textbf{The elevation of the wind base above the midplane $\phi_{\rm b}$.} The elevation of the base is largely controlled by the aspect ratio of the underlying disk \citep{Picogna_2021}, and accordingly scales similarly. A rough approximation (effectively assuming $M_*$-independent disk temperatures) is
    \begin{equation}
        \phi_{\rm b}(r;M_*) = 8.5^\circ \left(\frac{r}{\mathrm{au}}\right)^{1/4} \left(\frac{M_*}{M_{\odot}}\right)^{-0.5}.
    \end{equation}
    Given T Cha's stellar mass of $1.5\,M_\odot$, at 10s au, we expect $\phi_{\rm b} \gtrsim 14^\circ$ and so choose the $18^\circ$ models of \citet{Sellek_2021} for our analysis (the lowest $\phi_{\rm b}$ pre-calculated models fulfilling this criterion).
    \item \textbf{The launch angle of the wind with respect to the base $\chi_{\rm b}$}.} We assume $90^\circ$ since we expect large temperature jumps across the base, which will lead to strong acceleration normal to the base only and thus streamlines which emerge approximately perpendicular to the base).
\end{itemize}

For any parameter combination, the self-similar solutions predict a unique $\mathcal{M}_{\rm b}$ value \citep{Clarke_2016,Sellek_2021}, applicable to all streamlines (along with an associated velocity and density structure).
$b=1.5$, $\tau=0$, $\phi_{\rm b}=18^\circ$, \& $\chi_{\rm b}=90^\circ$ gives $\mathcal{M}_{\rm b}=0.437$.
Figure \ref{fig:Rz_maps} shows an example of a cross-section through the resulting wind model for a particular density normalization and inner radius; the number density of the gas is shown in the second panel, while the geometry of the resulting streamline is overlaid on the temperature profile (see Section \ref{sec:methods_radTrans}) in the first panel of Figure \ref{fig:Rz_maps}. Also indicated are the key angles defined above, as well as the opening angle at the base $\theta_{\rm b}=|90^\circ-\chi_{\rm b}-\phi_{\rm b}|=18^\circ$ and the maximum opening angle along the streamline, measured to be $\theta_{\rm max}=33^\circ$.

The density normalization is the most complicated thing to set.
It is connected to the mass-loss rate which is uncertain by several orders of magnitude between different models for photoevaporation where different parts of the high-energy spectrum drive the mass-loss.
Assuming $b<2$ \citep[as required for valid solutions,][]{Sellek_2021} and $r_{\rm in}<<r_{\rm out}$ then for a self-similar isothermal wind, the two are related through
\begin{equation}
    \dot{M}_{\rm nom} = \frac{4\pi\cos{\phi_{\rm b}}}{2-b}~\mathcal{M}_{\rm b}~\frac{G^{b} M_*^{b}}{c_{S}^{2b-1}}~r_{\rm out}^{2-b}~\rho_G
    \label{eq:Mdot_rhoG}
    ,
\end{equation}
where $\rho_G$ is the mass density at the gravitational radius and we assume $c_{S}=10\,\mathrm{km\,s^{-1}}$.
However since $b<2$, this expression diverges to large $r_{\rm out}$ so the density is sensitive to not only the assumed $c_{S}$ but also $r_{\rm out}$.

We define our models over a spherical region with an outer radius of $r_{\rm out}$ equal to the \textsuperscript{12}CO disk outer radius of $220\,\mathrm{au}$ \citep{Huelamo_2015}. This is the most optimistic scenario for an extended wind. Since photons propagate predominantly outwards, the inner regions are not affected by the outer parts of a wind, and less extended winds can be explored simply by cropping the outer radius during the analysis (noting as above that this affects the mass-loss rate).

The number density $n$ of the gas (Figure \ref{fig:Rz_maps}, second panel) is then derived from its mass density $\rho$ assuming $\rho = \mu m_{\rm H} n$ with $\mu=1.298$ for a gas of the composition we use in our radiative transfer simulations (see Section \ref{sec:methods_radTrans}). For our assumed parameters, this gives a number density at $r_G$ scaling as
\begin{equation}
    n_G = 1.6 \times 10^3\,\mathrm{cm^{-3}} \frac{\dot{M}_{\rm nom}}{10^{-10}\,M_{\sun}\,\mathrm{yr^{-1}}}
    \label{eq:nG}
    .
\end{equation}
For comparison, we list the critical densities for excitation of the lines of interest by electrons and neutral H at $7700\,\mathrm{K}$ in Table \ref{tab:ncrit}; contours showing where the wind attains these values are overlaid on the second and third panels in Figure \ref{fig:Rz_maps} respectively. However, note that although we provide values here, \textsc{mocassin} does not yet include neutral H as a collider, as the excitation rates have only just been calculated \citep{Yan_2024}; we explore the impact of this in Appendix \ref{appendix:collisions}.

\begin{table}[t]
    \centering
    \caption{Critical densities for excitation of noble gas lines by electrons and neutral H at 7700 K, as used in Figure \ref{fig:Rz_maps}.}
    \label{tab:ncrit}
    \begin{tabular}{cccc}
        \tableline
        Ion & Wavelength  & \multicolumn{2}{c}{Critical density / $\mathrm{cm^{-3}}$} \\
        &&  of electrons & of neutral H\tablenotemark{a}\\
        \tableline
        \ion{Ne}{2} & 12.81 & $6.2 \times 10^5$\tablenotemark{b} & $1.1 \times 10^7$ \\
        \ion{Ne}{3} & 15.55 & $2.4 \times 10^5$\tablenotemark{c} & $2.5 \times 10^6$ \\
        \ion{Ar}{2} &  6.98 & $3.8 \times 10^5$\tablenotemark{d} & $6.0 \times 10^7$ \\
        \ion{Ar}{3} &  8.99 & $3.0 \times 10^5$\tablenotemark{e} & $1.6 \times 10^7$ \\
        \tableline
    \end{tabular}
    \tablerefs{\textsuperscript{a}\citet{Yan_2024}, \textsuperscript{b}\citet{Saraph_1994}, \textsuperscript{c}\citet{McLaughlin_2000}, \textsuperscript{d}\citet{Pelan_1995}, \textsuperscript{e}\citet{Galvanis_1995}}
\end{table}

\subsubsection{Column density and the inner wind}
The absorption of photons at small radii can affect the outer disk \citep[for example in the scenario envisaged by][where an inner wind shields the outer disk until late in its evolution]{Pascucci_2020,Pascucci_2023}. We must therefore explore the effect of the inner parts of the wind.

The column density of gas in the self-similar models, which will be absorbing and attenuating the photons, may be calculated as
\begin{align}
    N(\phi) &= \int_{r_{\rm in}}^{r_{\rm out}} n_G \tilde{n}(\phi) \left(\frac{r}{r_G}\right)^{-b} dr \\
    &\propto n_G (r_{\rm out}^{1-b} - r_{\rm in}^{1-b})
    \label{eq:N_rin_rout}
    .
\end{align}
For $b>1$, this expression is dominated by small radii near $r_{\rm in}$, while for $b<1$, the column density is mostly accumulated at large radii.
Substituting equation \ref{eq:Mdot_rhoG} and assuming our fiducial $b=1.5$, then along the wind base:
\begin{equation}
    N = \frac{1.1\times10^{20}\,\mathrm{cm^{-2}}}{\mathcal{M}_{\rm b} \cos(\phi_{\rm b})} \left(\frac{\dot{M}}{10^{-8}\,M_{\sun}\,\mathrm{yr^{-1}}}\right) 
    \left(\frac{r_{\rm in}}{1\,\mathrm{au}}\right)^{-1/2}
    \label{eq:N_values}
    .
\end{equation}
Hence we can see that the two main parameters determining the optical depth in our models will be the inner radius and the mass-loss rate (the outer radius enters implicitly through the mass-loss rate e.g. Equation \ref{eq:Mdot_rhoG}).

To contextualize these values (see also the cross-sections in Figure \ref{fig:crossSec}), note that the penetration depth of EUV photons in neutral hydrogen is typically $10^{17}-10^{19}\,\mathrm{cm^{-2}}$, hence as expected from the mass-loss rate of EUV-driven winds, the winds will be optically thin for $\dot{M}\lesssim 10^{-9}\,M_\odot\,\mathrm{yr^{-1}}$.
On the other hand $1\,\mathrm{keV}$ X-rays - which may perform K shell photoionization of Ne - can penetrate a neutral hydrogen column of $10^{22}\,\mathrm{cm^{-2}}$ and the winds are generally optically thin to these even for the highest density normalization we consider. This is also to be expected since \citet{Sellek_2022} found that the most effective X-rays for wind driving (which are distinct from the most effective at photoionizing Ne) are those for which the corresponding wind presents an optical depth at that energy of $\tau \approx 1$, and that this should occur around $500\,\mathrm{eV}$ for typical X-ray luminosities.

In order to vary the absorbing column, we thus vary the inner radius $r_{\rm in}$ of the grid on which we perform radiative transfer.
Thus, we are defining $r_{\rm in} \in {0.03, 0.1, 0.3, 1.0, 3.0}\,r_G$ as the radius inside of which we set the gas density to be zero.
Such $r_{\rm in}$ values are motivated by the fact that in the photoevaporative scenario, the wind ceases being launched on a scale somewhere in the range $0.1-1\,r_G$, smoothly transitioning to a hydrostatic atmosphere inside of this point. An MHD disk wind may carry on inside this point, so we also employ a smaller value of $0.03\,r_G$ to crudely explore this scenario, while $3\,r_G$ allows us to consider a case where the wind only launches outside of the mm cavity which is $\sim30\,\mathrm{au}$ in radius \citep{Francis_2020}.

However, we emphasize that in general, since we do not model the underlying disk (only the wind region), these choices of radius do not directly constrain or reflect the presence or location of a gas cavity or gap in the disk. That is to say, the gas may extend in closer to the star than $r_{\rm in}$ at the midplane while stopping approximately at $r_{\rm in}$ at greater elevations, and not affect our results in any way. While this cut off \textit{may} result from the absence of gas in the disk to feed the wind inwards of some cavity radius, it can also arise as result of the limited radii at which a wind can lift significant additional material to higher elevations. In reality, the edge of the wind is not likely sharp as modeled here but smoothly transitions to a hydrostatic inner atmosphere; what matters most for our results is where the column density is mostly accumulated. Therefore, only if this happens in the wind will $r_{\rm in}$ necessarily be a good measure of the wind's true inner radius.

\subsubsection{Defining mass-loss rates}
\label{sec:methods_Mdot}
As discussed, estimates of the mass-loss rate between photoevaporation models vary considerably. One key factor in this is the driving radiation: since X-rays are more penetrating than EUV they are expected to drive much denser winds with mass-loss rates $1-2$ orders of magnitude than EUV. For example, given the $<4.1\times 10^{41}\,\mathrm{s^{-1}}$ EUV photon flux of T Cha \citep{Pascucci_2014} and its $1.5\,M_{\sun}$ mass, the expected EUV photoevaporation rate \citep{Shu_1993} would be $\dot{M}_{\rm PE} \lesssim 1.2\times10^{-9}\,\mathrm{yr^{-1}}$, while in the other extreme, the X-ray photoevaporation rate predicted by \citet{Picogna_2021} for a star of this mass is $\dot{M}_{\rm PE} \sim 5.9\times10^{-8}\,\mathrm{yr^{-1}}$.
We use the outer gas radius of $220\,\mathrm{au}$ \citep{Huelamo_2015} for $r_{\rm out}$ in Equation \ref{eq:Mdot_rhoG}, 
in order to scale $\rho_G$ so as to produce nominal mass-loss rates $\dot{M}_{\rm nom} \in {10^{-10},10^{-9},10^{-8},10^{-7}}\,M_{\sun}\,\mathrm{yr^{-1}}$ that span this large range of possible values. 

To remedy the issue of the mass-loss rate diverging to large radius in the self-similar models, we note that the line emissivity in the outer regions of the wind, which will be below the critical density, scales as $n^2$ and therefore diverges more slowly, if at all.
Thus from now on we will quote as the ``true'' mass-loss rate the ``observable mass-loss rate'' for which we would have evidence from the [\ion{Ne}{2}] line (this being our most confidently detected line) with MIRI-MRS given the absolute flux error of 5 per cent \citep{Argyriou_2023}, which dominates over the $1\sigma$ statistical uncertainty \citep{Bajaj_2024}. 

\begin{figure}[t]
    \centering
    \includegraphics[width=\linewidth]{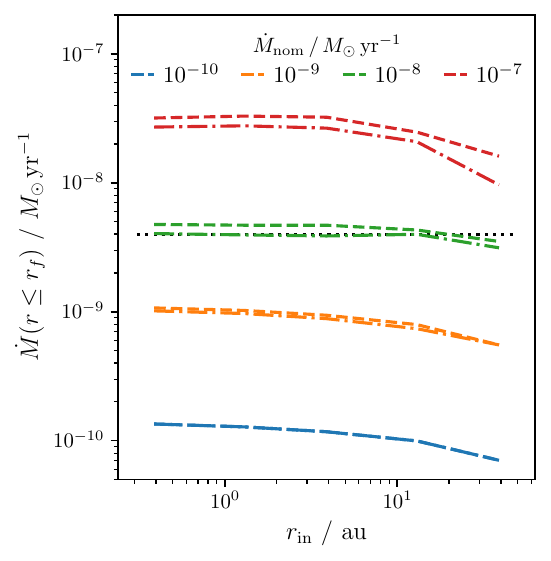}
    \caption{Observable mass-loss rates inferred from our hiLX\_TCha\_sUV models. The models were constructed with nominal mass-loss normalizations of $10^{-10}\,M_{\sun}\,\mathrm{yr^{-1}}$ (blue), $10^{-9}\,M_{\sun}\,\mathrm{yr^{-1}}$ (orange),  $10^{-8}\,M_{\sun}\,\mathrm{yr^{-1}}$ (green), and $10^{-7}\,M_{\sun}\,\mathrm{yr^{-1}}$ (red); the observable rates were determined by taking the outer radius of the wind to be the radius containing $95$ (dot-dashed) and $97$ (dashed) per cent of the [\ion{Ne}{2}] 12.81 $\mu$m emission. The black dotted line indicates T Cha's observed accretion rate  \citep{Cahill_2019} for comparison.}
    \label{fig:true_Mdot}
\end{figure}

To obtain the observable mass-loss rate, we calculate the observed outer radius as the spherical radius $r_f$ containing a fraction $f = 0.95$ of the line luminosity; roughly speaking (assuming the model produces the correct total [\ion{Ne}{2}] luminosity) this is designed to account for the fact that the outer reaches of the wind could not be confidently detected at a statistical level. 
For this purpose we also normalize the velocities of the self-similar model to the sound speeds derived from the temperature maps provided by the radiative transfer (even though these are not strictly isothermal).
Depending on the assumed inner radius and the irradiating spectrum, this results in observable mass-loss rates for each nominal normalization in the ranges given in Table \ref{tab:Mdotnom_to_Mdotobs}.\footnote{For the lowest $\dot{M}_{\rm nom}$, the sound speed can slightly exceed the $10\,\mathrm{km\,s^{-1}}$ assumed in Equation \ref{eq:Mdot_rhoG}, allowing the ``observable'' mass-loss rate to be slightly higher than the ``nominal'' one.}
As an example, in Figure \ref{fig:true_Mdot} we show the calculated $\dot{M}$ as a function of $r_{\rm in}$ for one of our model sets hiLX\_TCha\_sUV (see \ref{sec:methods_radTrans} for details).
The figure also includes equivalent values for $f=0.97$ (equivalent to the $5 \sigma$ level using the statistical uncertainty), showing that the results are not hugely sensitive to the choice of $f$. The ranges quoted in Table \ref{tab:Mdotnom_to_Mdotobs} and shown in Figure \ref{fig:true_Mdot} are both narrow enough and distinct enough to use as labels henceforth.

\begin{table}[t]
    \centering
    \caption{The observable mass-loss rates derived for each nominal mass-loss rate normalization.}
    \label{tab:Mdotnom_to_Mdotobs}
    \begin{tabular}{cc}
         \tableline
         $\dot{M}_{\rm nom}\,/\,M_{\sun}\,\mathrm{yr^{-1}}$ & $\dot{M}\,/\,M_{\sun}\,\mathrm{yr^{-1}}$\\
         \tableline
            $10^{-10}$  &   $0.7-2 \times10^{-10} M_{\sun}\,\mathrm{yr^{-1}}$ \\
            $10^{-9}$   &   $0.5-1 \times10^{-9}  M_{\sun}\,\mathrm{yr^{-1}}$ \\
            $10^{-7}$   &   $3-5   \times10^{-9}  M_{\sun}\,\mathrm{yr^{-1}}$ \\
            $10^{-7}$   &   $1-3   \times10^{-8}  M_{\sun}\,\mathrm{yr^{-1}}$ \\
         \tableline
    \end{tabular}
\end{table}

\subsection{Spectra and radiative transfer}
\label{sec:methods_radTrans}
We use the Monte Carlo photoionization radiative transfer code \textsc{mocassin} \citep{Ercolano_2003,Ercolano_2005,Ercolano_2008} to calculate the temperature and ionization equilibrium at each point the wind. 
As well as the density distributions described above, the other key input to these radiative transfer simulations is one of five high-energy spectra. Details of each spectrum, which were produced using \textsc{pintofale} \citep{Kashyap_2000}, are given in Table \ref{tab:spectra_NeII} and summarized below.

\begin{table*}[t]
    \centering
    \caption{Properties of Different Spectra Used}
    \label{tab:spectra_NeII}
    \setlength\tabcolsep{2pt}
    \begin{tabular}{ccccccc}
         \tableline
         Name                  & Components & Elemental & $L_{\rm X}$ & $\Phi_{\rm EUV}$ & $L_{\rm tot}$ & Salient \\
         && Abundance & ($\mathrm{erg~s^{-1}}$) & ($\mathrm{s^{-1}}$) & ($\mathrm{erg~s^{-1}}$) & Features \\
         && Template & [Energy Range &&& \\         
         &&& (keV)] &&& \\
         \tableline

         Ercolano\_RSCVn
         & RS CVn template & Solar & $2.7\times10^{30}$ & $1.2\times10^{41}$ & $1.06\times10^{31}$ & Used by \\
         & (peaks at $T=1.8\times10^7\,\mathrm{K}$ & Photospheric & [$0.3-10$] &&& \citet{Ercolano_2010} \\
         & and $T=10^4\,\mathrm{K}$) & (5) & (2) &&& \citet{Ercolano_2016}; \\
         & (1) &&&&& Lowest $L_{\rm X}$ \\
         \tableline
         
         loLX\_TCha
         & $T_1 = 0.9 \times 10^7\,\mathrm{K}$ & $0.8\times$ Solar & $4.4\times10^{30}$ & $7.0\times10^{39}$ & $5.23\times10^{30}$ & Lowest EUV\\
         & $T_2 = 2.7 \times 10^7\,\mathrm{K}$ & Photospheric & [$0.15-8$] && \\
         & (3) & (6) & (3) && \\
         & $EM_2/EM_1 = 3$ &&&& \\
         \tableline

         loLX\_TCha\_sUV
         & " & " & " & $4.1\times10^{41}$ & $1.58\times10^{31}$ \\
         & $T_3 = 10^4\,\mathrm{K}$ &&& (4) & \\
         & $EM_3/EM_1 = 35.1 $ &&&& \\
         \tableline

         hiLX\_TCha
         & $T_1 = 0.35 \times 10^7\,\mathrm{K}$ & TW Hya & $3.7\times10^{31}$ & $8.6\times10^{40}$ & $4.52\times10^{31}$ & Highest $L_{\rm X}$ \\
         & $T_2 = 2.1 \times 10^7\,\mathrm{K}$ & X-ray plasma & [$0.15-8$] && \\
         & (3) & Model C of (7) & (3) && \\
         & $EM_2/EM_1 = 10$ &&&& \\
         \tableline
         
         hiLX\_TCha\_sUV
         & " & " & " & $4.1\times10^{41}$ & $5.38\times10^{31}$ & Highest $L_{\rm X}$  \\
         & $T_3 = 10^4\,\mathrm{K}$ &&& (4) && and EUV \\
         & $EM_3/EM_1 = 7.3$ &&&& \\
         \tableline
    \end{tabular}
    \tablerefs{(1) \citet{Ercolano_2009} (2) \citet{Gudel_2010}, (3) \citet{Sacco_2014}, (4) \citet{Pascucci_2014} (5) \citet{Grevesse_1998} (6) \citet{Grevesse_1992} (7) \citet{Brickhouse_2010}}
    \tablecomments{$L_{\rm X}$ values are rescaled from the original references based on distance from Gaia DR3. Since (3) do not provide values, $EM_2/EM_1$ is estimated from the plots of \citet{Preibisch_2005} using the inferred $L_{\rm X}$. $EM_3/EM_1$ was adjusted such that the $\Phi_{\rm EUV}$ matched the upper limit of \citet{Pascucci_2014}.}
\end{table*}

Firstly, we test a spectrum we call ``Ercolano\_RSCVn'' - which is the \citet{Ercolano_2009} model FS0H2Lx1 - to enable direct comparisons with \citet{Ercolano_2010,Ercolano_2016}.
This uses RS CVn binaries (which contain a giant with comparable deep convective zones to T Tauri stars) as a template for a spectrum containing both EUV and X-ray components. We normalize this spectrum to the observed X-ray luminosity of $2.7\times10^{30}\,\mathrm{erg\,s^{-1}}$ \citep{Gudel_2010}; this results in an EUV flux of $1.2\times10^{41}\,\mathrm{s^{-1}}$, consistent with the \citet{Pascucci_2014} upper limit of $4.1\times10^{41}\,\mathrm{s^{-1}}$.

For a more direct model of T Cha's X-ray emission, we turn to \citet{Sacco_2014} who provide two-temperature fits.
These were performed for two different abundance patterns, resulting in two possible spectra: a harder, less luminous spectrum results from solar abundances, while a softer, more luminous spectrum results from assuming a pattern similar to TW Hya \citep{Brickhouse_2010}.
We find the most important differences in our results are due to the luminosities, hence we refer to these as ``loLX\_TCha'' and ``hiLX\_TCha'' respectively \citep[although both have a higher $L_{\rm X}$ than][]{Gudel_2010}.
Moreover, since as both of these spectra are comfortably below the \citet{Pascucci_2014} upper limit on the EUV flux, we create an additional version of each (denoted ``loLX\_TCha\_sUV'' and ``hiLX\_TCha\_sUV'') which include an additional $10^4\,\mathrm{K}$ soft EUV component scaled to match this upper limit. Thus we can span the possible contributions from EUV.

For the abundance of each element relative to hydrogen in the wind, we follow typical values in the literature \citep[e.g.][]{Ercolano_2009,Hollenbach_2009b,Wang_2017} which are usually based on the depleted gas phase ISM abundances of \citet{Savage_1996}. These are detailed in Table \ref{tab:abundances}.

\begin{table}[t]
    \centering
    \caption{Number abundances (with respect to H) of the elements included in the photoionization radiative transfer}
    \label{tab:abundances}
    \begin{tabular}{cc}
         \tableline
         Element & Abundance\\
         \tableline
            He  &   0.1 \\
            C   &   $1.4\times10^{-4}$ \\
            N   &   $8.32\times10^{-5}$ \\
            O   &   $3.2\times10^{-4}$ \\
            Ne  &   $1.2\times10^{-4}$ \\
            Mg  &   $1.1\times10^{-6}$ \\
            Si  &   $1.7\times10^{-6}$ \\
            S   &   $2.8\times10^{-5}$ \\
            Ar  &   $6.3\times10^{-6}$ \\
            Fe  &   $1.7\times10^{-7}$ \\
         \tableline
    \end{tabular}
\end{table}

\begin{figure}
    \centering
    \includegraphics[width=\linewidth]{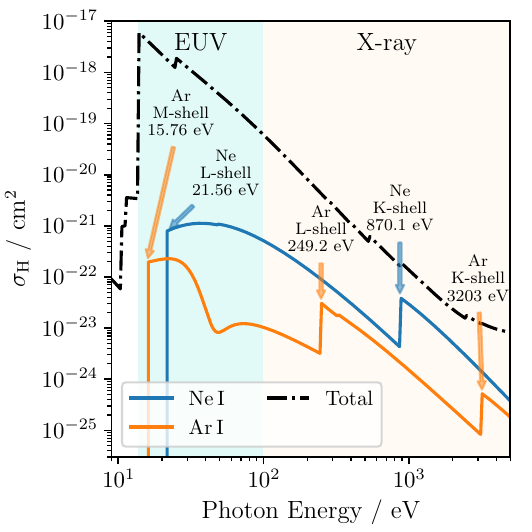}
    \caption{The photoionization cross-sections (per hydrogen atom) $\sigma_{\rm H}$ used in this work. The black dot-dashed line shows the total cross-section of a neutral gas while the blue and orange lines indicate that of \ion{Ne}{1} and \ion{Ar}{1} respectively, with major thresholds corresponding to ionization from different electron shells labelled. The typical energy ranges described as EUV ($13.6-100\,\mathrm{eV}$) or X-ray ($>100\,\mathrm{eV}$) are indicated by the shaded zones.}
    \label{fig:crossSec}
\end{figure}

\textsc{mocassin} uses the photoionization cross-sections from \citet{Verner_1996} for the outer (valence) shell ionization and \citet{Verner_1995} for the inner shell ionization (which becomes relevant at X-ray energies).
Figure \ref{fig:crossSec} shows the total photoionization cross-section for a neutral gas of this composition, along with those of Ne and Ar with their major ionization thresholds indicated. 
The charge on an ion may be lowered either by charge exchange, for which \textsc{mocassin} uses the values tabulated by \citet{Kingdon_1996}, or recombination, for which \textsc{mocassin} mostly follows (including for our species of interest) \citet{Badnell_2006a,Badnell_2006b}.

For each combination of density grid and spectrum, we run \textsc{mocassin} \citep{Ercolano_2003,Ercolano_2005,Ercolano_2008} for eight iterations with $10^8$ photon packets followed by a final iteration with $10^{10}$ photon packets to reduce noisiness in the ionization state. The outputs include gas temperatures (Figure \ref{fig:Rz_maps}, first panel), electron densities (Figure \ref{fig:Rz_maps}, third panel) and emissivities of selected lines of interest in each cell (Figure \ref{fig:Rz_maps}, second row).

\begin{figure*}
    \centering
    \includegraphics{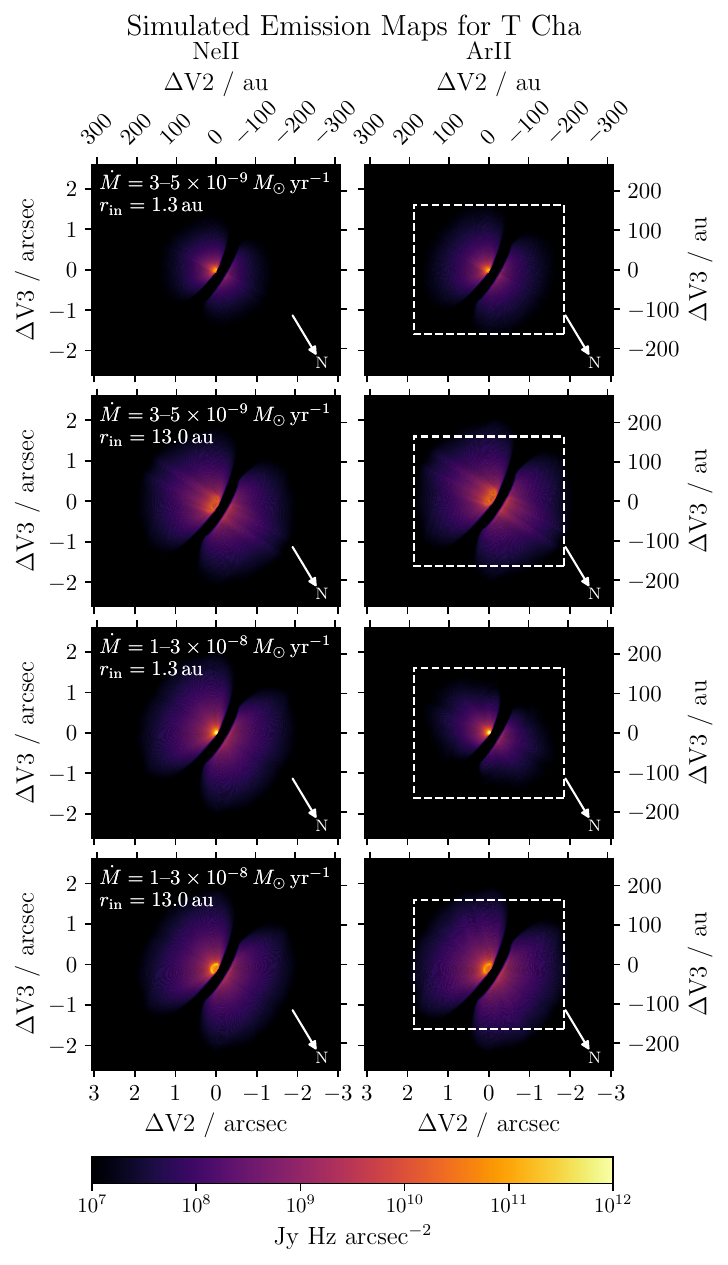}
    \caption{Examples of the integrated emission maps of [\ion{Ne}{2}] (left) and [\ion{Ar}{2}] (right) for different mass-loss rates and inner radii for models using the hiLX\_TCha\_sUV spectrum. These are provided as inputs to \textsc{MIRISim} for the synthetic imaging. Each of the left-hand panels shows the full field of view in Channel 3 (which contains the [\ion{Ne}{2}] line); the right-hand panels show the same FOV for ease of comparison between the two lines with the FOV of Channel 1 (which contains the [\ion{Ar}{2}] line) shown by the dashed box.}
    \label{fig:emission_maps}
\end{figure*}

In Figure \ref{fig:emission_maps} we show examples of the [\ion{Ne}{2}] and [\ion{Ar}{2}] emission maps from \textsc{mocassin} projected to the inclination and position angle (on the MIRI-MRS detector) of T Cha and integrated both along the line of sight and in wavelength. All the images show diffuse, often extended, emission. Two other sorts of features can be seen. In the bottom row, a ring of emission at $r_{\rm in}$ - which is bright due to the additional ionization from soft EUV incident on the inner edge of the wind - can be distingiushed when the inner radius is large enough that it can dominate the flux (see also Figure \ref{fig:Rz_maps_in0}). In the second row, distinctive conical shapes are seen which trace the limb-brightened ionization front inside of which the [\ion{Ne}{2}] emission is weak since most of the Ne is in \ion{Ne}{3} (see also Figure \ref{fig:Rz_maps_Mdot8}). This coincides with the $\tau=1$ surfaces for 20-40 eV photons near the L-shell ionization thresholds of Ne, meaning this wind model is optically thin to EUV at high altitudes but optically thick at the wind base.

\subsection{Calculation of line fluxes and profiles}
The most simple product from the radiative transfer simulations is the total luminosity of each emission line.
We use the fluxes directly from MOCASSIN, rather than from the synthetic images that we later create. This is because despite our best efforts (Section \ref{appendix:MIRISim}), we were not always able to reliably recover the total input line (or continuum) flux from the synthetic spectrum.
To calculate the visible fluxes, we simply integrate over the volume of the wind, including only cells which we consider visible to the observer.
For this purpose we assume that the midplane of the disk is infinitely optically thick (with all other material being optically thin) inside the $R_{\rm out}=220\,\mathrm{au}$ gas disk \citep{Huelamo_2015}, except for possibly inside a dust cavity of $R_{\rm cav}=15\,\mathrm{au}$ \citep{Xie_2023}. Thus, any cell in the wind on the far side of the disk is only included if the line of sight does not cross the optically thick midplane. Mathematically this means we mask out cells where $z<0$ and
\begin{equation}
    R_{\rm cav}^2 < (R \cos(\phi) - z \tan(i))^2 + (R\sin(\phi))^2 < R_{\rm out}^2
    \label{eq:cavity_mask}
    .
\end{equation}

To obtain the line shape, in each cell we calculate a thermally-broadened Gaussian centered on the line of sight velocity $v_{\rm los}$ \citep[e.g.][]{Rybicki_1979}
\begin{equation}
    L(\rm v;\mathbf{r}) = \frac{1}{\sqrt{2\pi}v_{\rm th}} \exp\left( - \frac{(v-v_{\rm los})^2}{2v_{\rm th}^2} \right) L(\mathbf{r})
    ,
    \label{eq:broadenedv}
\end{equation}
where $v_{\rm th} = \sqrt{m_H/m_i} c_{\rm S}$ (for isothermal sound speed $c_{\rm S}$) and
\begin{align}
    v_{\rm los} =& (- \sin(\theta)\cos(\phi)\sin(i) - \cos(\theta)\cos(i) ) v_r \\
            &+ (- \cos(\theta)\cos(\phi)\sin(i) + \sin(\theta)\cos(i) ) v_\theta \nonumber \\
            &+ \sin(\phi)\sin(i) v_\phi \nonumber
            .
    \label{eq:vlos}
\end{align}
$v_r$ and $v_\theta$ are obtained from the self-similar models by normalizing to the local $c_{\rm S}$ calculated from the \textsc{mocassin} temperatures. We assume the azimuthal component equals the Keplerian velocity $v_\phi = v_{\rm K}$ at the base and conserve angular momentum along each streamline.

The contribution of all visible cells to the line profile is then summed and the resulting profile convolved with a Gaussian of width $10\,\mathrm{km~s^{-1}}$ in order to degrade the profile to the typical $R=30000$ spectral resolution of high resolution [\ion{Ne}{2}] spectra \citep{Pascucci_2020}. Somewhat higher resolution [\ion{O}{1}] spectra are available so for comparison to those we assume $R=45000$.

\subsection{Synthetic imaging}
\label{sec:methods_synthetic}
To produce synthetic JWST observations we use the \textsc{MIRISim} python package Version 2.4.2 \citep{Klaassen_2021}. \textsc{MIRISim} allows a ``scene'' to be built from several components. It then simulates an observation of this scene by the MIRI instrument of choice suitable for input into the pipeline: in our case detector images from the Medium Resolution Spectrometer. In so doing, it accounts for several distortions and transformations due to both the optics and detectors.
A few features were missing or needed correcting or updating in \textsc{MIRISim}. We briefly summarise the modifications we made in Appendix \ref{appendix:MIRISim}.
We note that \textsc{MIRISim} does not yet model the contribution of stray light (so we disable this step when reducing the data) and that there is a known inconsistency with the handling of reference pixels as of Version 2.4.2.

The simulation settings such as the dither pattern, number of integrations and groups per integration were kept the same as the observational settings \citep{Pascucci_2021,Bajaj_2024} such that the spatial sampling and exposure are comparable to the real data.

\subsubsection{Scenes}
Our scenes are built from the following components:
\begin{itemize}
    \item \textbf{A low-level uniform background representing the telescope thermal background \citep[following][]{Glasse_2015} and zodiacal light.}
    \item \textbf{The continuum emission from the disk and star.}
    A point source with a $1.36~L_\odot$, $5400\,\mathrm{K}$ blackbody represents the star \citep{Olofsson_2011}.
    To this we add the disk continuum emission, which \textsc{MIRISim} models with a Sérsic profile.
    Since most of the disk's dust will be at temperatures cold enough that the black-body peak lies at much longer wavelengths than those of observation, we may assume that we may use the Wien approximation for the dust emission. In this case, the emission scales proportional to $\exp(-\frac{hc}{\lambda  k_B T})$. Assuming $T\propto R^{-q}$, this results in an exponent proportional to $R^q$. Hence, we may identify the Sérsic index $n=1/q \approx 2$ for the typical temperature profile of a disk.
    Based on the moderately resolved continuum \citep{Bajaj_2024}, we provide this profile with a half-light radius of $0.33\,\mathrm{arcsec}$.
    We choose $R_{\rm in}=15\,\mathrm{au}$ based on SED modeling of the observed JWST spectrum, Xie et al. in prep. and do not assume an outer truncation radius, since it would anyway likely be in the tail of the tapered profile.
    The orientation of the disk has been measured using 3 mm continuum emission - which traces large dust grains that have settled to the midplane - to be an inclination of $i=73^\circ$ and position angle $PA=113^\circ$ \citep{Hendler_2018}.
    However \citet{Huelamo_2015} found a slightly lower value of $i=67^\circ$ by performing a fit to the gas emission, which has a greater vertical extent than the large, settled, dust grains. 
    Since the small dust grains that produce the micron continuum emission ought to follow the vertical distribution of the gas we therefore use $q=\cos(67^\circ)$ for the aspect ratio of our model disk.
    Moreover, \textsc{MIRISim} assumes a fixed position angle of $PA_{\rm V3}=0^\circ$ for the orientation of the JWST field of view on the sky. Therefore, to ensure our disk has the same orientation on the detector as the observations, we offset its position angle by the actual $PA_{\rm V3}=145^\circ$ of the observation such that $PA_{\rm simulated}=113^\circ-145^\circ=-32^\circ$.
    Finally, for the SED of the disk we use a fit to the observed continuum spectrum \citep{Bajaj_2024}.
    \item \textbf{The line emission from the wind}.
    For each line we provide \textsc{MIRISim} with a FITS file containing the $L(v;\mathbf{r})$ (equation \ref{eq:broadenedv}, Figure \ref{fig:emission_maps}), integrated along the line of sight and where the velocities are mapped onto wavelengths.
    We select for this purpose our best set of models with line luminosities not too far from the observed ones. However to control for exposure and sensitivity effects we then scale the emission in each line to the observed luminosity after accounting for which regions of the wind are visible.
    For lines of sight that intersect the optically thick dust disk, only the nearside wind contributes. To work out where the disk is optically thick, we use the SED modeling by \citet{Xie_2023}. This suggests that the optical depth at $50\,\mathrm{au}$ at $12.81\,\mu\mathrm{m}$ would be $1.6$ if viewed face on. Since the disk is thin $H/R \tan(i)<1$ \citep[$H/R=0.1$][]{Xie_2023}, we adopt the usual approximation that the optical depth along the line of site is boosted by a factor $1/\cos(i)$. \citet{Xie_2023} fit a $1/R$ surface density profile with exponential cut-off, with $R_C>50\,\mathrm{au}$ \citep{Pohl_2017,Huelamo_2015}. Therefore we estimate the disk's optical depth as
    \begin{equation}
        \tau \approx \frac{1.6}{\cos(i)} \left(\frac{R}{50\,\mathrm{au}}\right)^{-1} \exp\left(- \frac{R-50\,\mathrm{au}}{R_C}\right)
        ,
    \end{equation}
    Assuming the lower limit of $R_C=50\,\mathrm{au}$, the disk rapidly becomes optically thin in the exponential tail at $R_{\tau=1}\approx100\,\mathrm{au}$; for simplicity we therefore assume the disk is optically thick between $R_{\rm cav}=0\,\mathrm{au} \to R_{\rm out}=100\,\mathrm{au}$ (equation \ref{eq:cavity_mask}), with the receding wind being visible at larger radii. We also test otherwise identical models where $R_{\rm cav}=15\,\mathrm{au}$ \citep{Xie_2023}.
\end{itemize}

We likewise create a set of synthetic observations with only the thermal background component to use to conduct background subtraction.

We also create synthetic observations corresponding to the standard star HD37962 and its background \citep[Program CAL/CROSS 1538][]{Gordon_2019} such that we may compare the synthetic T Cha observations to the telescope PSF \citep[as per][]{Bajaj_2024} while ensuring a consistent PSF model is used.

\subsubsection{Reduction}
We reduce these synthetic observations in the same way as the real observations using version `1.11.2' of the JWST pipeline \citep{JWST_pipeline} using the `ifualign' mode to stay in the detector plane. The background is subtracted on a pixel-by-pixel basis after the cube creation, followed by a spaxel-by-spaxel continuum subtraction to generate a map of the line emission only. For full details see \citet{Bajaj_2024}.

\section{Constraints from line luminosities and ratios}
\label{sec:ratios}

In this section, we compare the luminosities (and line ratios) of the mid-IR emission lines predicted directly from the \textsc{mocassin} radiative transfer to those derived from observations.
While we focus on T Cha, we include for comparison other sources in the literature with [\ion{Ne}{2}] emission which has been studied with high-resolution spectroscopy such that it could be classified as an LVC, and which have detections or upper limits for [\ion{Ne}{3}] and/or [\ion{Ar}{2}] derived in the same work as a [\ion{Ne}{2}] detection.
As well as T Cha, this yielded a further 3 sources from \citet{Pascucci_2020} - V4046 Sgr, TW Hya and CS Cha - as well as RX J1615.3-3255 \citep{Sacco_2012}.
As a result of selecting for [\ion{Ne}{2}] LVCs, these sources all show signs of being more evolved \citep{Pascucci_2020}. Their infrared SEDs show signs of dust depletion in the inner regions leading them to be classified as transition disks and their accretion rates are all below $10^{-8}<M_\odot\,\mathrm{yr^{-1}}$. While this means the sources are not representative of the population of disks as a whole, limiting our discussion to this selection means that we avoid trying to compare photoevaporative wind models to objects where they would not be a valid description, for example systems where emission more likely originates in a high-velocity jet \citep[where shock heating and collisional ionization are important][]{Hollenbach_1989}. As such we exclude, for example, Sz 102 \citep{Lahuis_2007} - where velocity-resolved optical lines of [\ion{Ne}{3}] indeed suggest an origin in jets \citep{Liu_2014} - and SZ Cha \citep[][see discussion]{Espaillat_2013,Espaillat_2023}.
Other than the new JWST measurements for T Cha, these additional data come from Spitzer spectra. Details of these are given in Table \ref{tab:JWSTTCha}.

Note that for T Cha, the uncertainty is dominated by the $\sim5$ per cent absolute flux error, rather than the statistical $1\sigma$ errors listed in Table \ref{tab:JWSTTCha} (which are derived as the standard deviation on a Monte-Carlo sample of 10,000 Gaussian fits); the two should be added in quadrature \citep{Bajaj_2024} and thus is what we indicate in our error bars (though they are generally too small to be visible except in relation to the [\ion{Ar}{3}] flux). There is very little molecular emission in the T Cha spectrum, with only a single line of H\textsubscript{2} ($9.66\,\mu\mathrm{m}$) detected. The [\ion{Ne}{2}] line does sit on top of a PAH band but this is effectively removed by the continuum subtraction. Thus the line fluxes are not affected by any line blending and can safely be compared to our model which consists of continuum and Ne/Ar lines only.

\begin{table*}[t]
    \centering
    \caption{Literature measurements (including, where published, $3\sigma$ upper limits) of line luminosities for disks with [\ion{Ne}{2}] LVCs.}
    \label{tab:JWSTTCha}
    \begin{tabular}{ccccccc}
        \tableline
        Ion & Wavelength  & \multicolumn{5}{c}{Luminosity / $ 10^{27}\,\mathrm{erg\,s^{-1}}$} \\
        & $\mu\mathrm{m}$ &  T Cha & V4046 Sgr & TW Hya & CS Cha & RX J1615.3-3255\\
        \tableline
        & Distance / pc\tablenotemark{a}& 102.7 & 71.48 & 60.14 & 168.8 & 155.6 \\
        \tableline
        \ion{Ne}{2} & 12.81 & $63.1 \pm 0.1$\tablenotemark{b}   & $53.2 \pm 0.8$\tablenotemark{c}   & $24.1 \pm 0.7$\tablenotemark{d}  & $124 \pm 2$\tablenotemark{f}       & $80 \pm 10$\tablenotemark{e} \\
        \ion{Ne}{3} & 15.55 & $6.6 \pm 0.3$\tablenotemark{b}   & $<6.8$\tablenotemark{c}           & $1.1 \pm 0.8$\tablenotemark{d}  & $10 \pm 2$\tablenotemark{f}      & $<80$\tablenotemark{e} \\
        \ion{Ar}{2} &  6.98 & $54.5 \pm 0.2$\tablenotemark{b}   & -                          & $<36.2$\tablenotemark{e}          & $<101$\tablenotemark{e}                & $<220$\tablenotemark{e} \\
        \ion{Ar}{3} &  8.99 & $1.0 \pm 0.3$\tablenotemark{b} & - & - & - & - \\
        \tableline
    \end{tabular}
    \tablerefs{\textsuperscript{a}\citet{Gaia_2022}, \textsuperscript{b}\citet{Bajaj_2024}, \textsuperscript{c}\citet{Rapson_2015b}, \textsuperscript{d}\citet{Najita_2010}, \textsuperscript{e}\citet{Szulagyi_2012}, \textsuperscript{f}\citet{Espaillat_2013}}
\end{table*}

\subsection{Ionization constraints}
In order that initially we are not sensitive to overall scalings such as abundances, we start by comparing the predicted Ne ([\ion{Ne}{3}] $15.55~\mu\mathrm{m}$ to [\ion{Ne}{2}] $12.81~\mu\mathrm{m}$) and Ar ([\ion{Ar}{3}] $8.99~\mu\mathrm{m}$ to [\ion{Ar}{2}] $6.98~\mu\mathrm{m}$) line ratios from our models; we expect these quantities to provide information about the levels of ionization and EUV heating in the wind \citep{Hollenbach_2009b} as well as about the shape of the EUV/X-ray spectra.

\begin{figure*}[h]
    \centering
    \includegraphics[width=\linewidth]{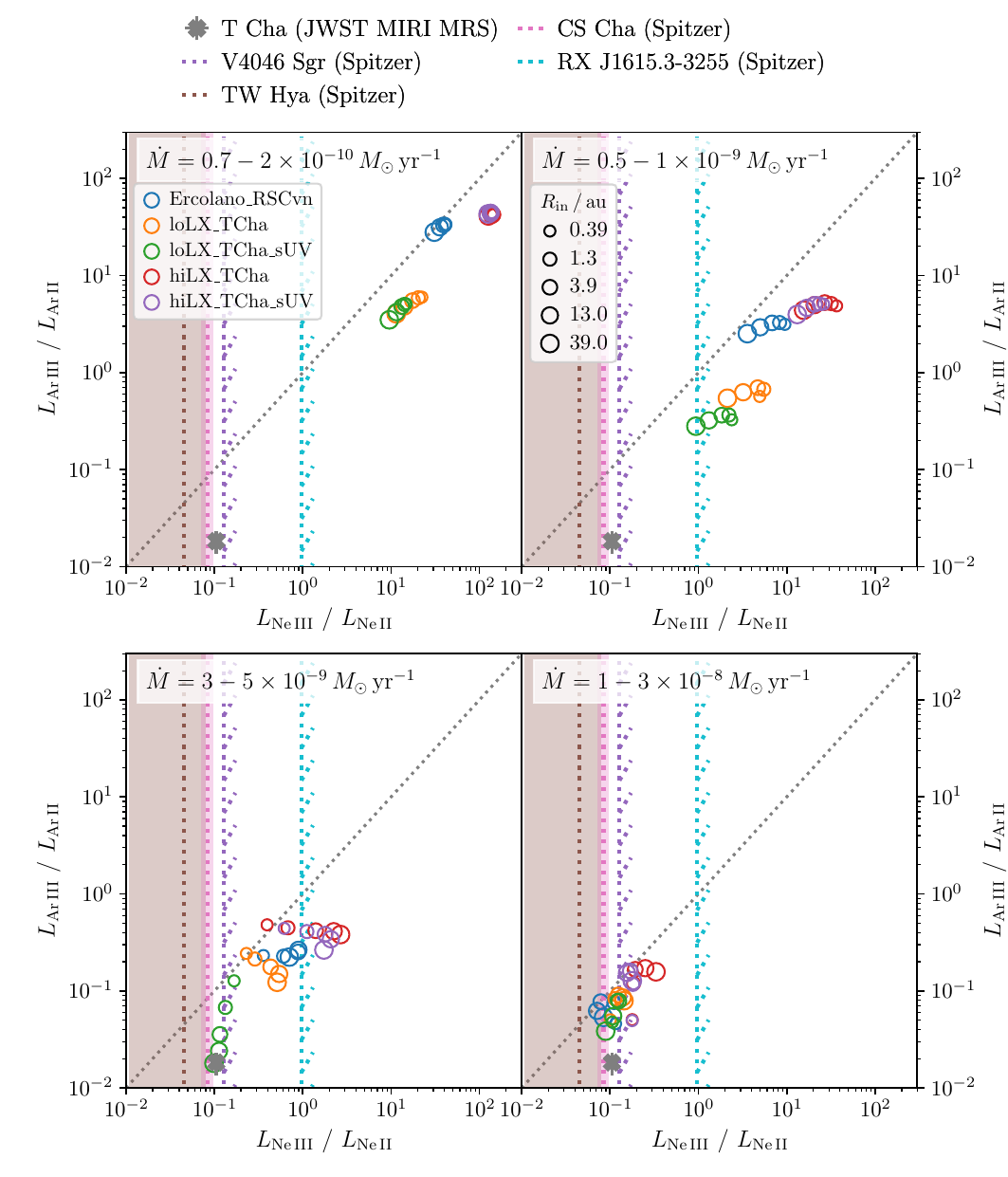}
    \caption{Comparison of the ratio of the [\ion{Ar}{3}] $8.99~\mu\mathrm{m}$ and [\ion{Ar}{2}] $6.98~\mu\mathrm{m}$ lines to the ratio of the [\ion{Ne}{3}] $15.55~\mu\mathrm{m}$ and [\ion{Ne}{2}] $12.81~\mu\mathrm{m}$. Each set of points of a given color represents a particular one of the five spectra considered in this work; increasing point sizes indicate the sequence of inner radii from smallest to largest. Results are shown for different $\dot{M}$ on each panel. JWST measurements for T Cha are shown as the gray cross in lower left - the error bar includes the 5 per cent absolute flux error added to the $1\sigma$ error in quadrature \citep{Bajaj_2024} but is generally too small to be visible except in relation to the [\ion{Ar}{3}] flux. The shaded regions indicate values that would be consistent with CS Cha (pink) or TW Hya (brown),  while the hashed vertical lines indicate the regions excluded by the upper limits on [\ion{Ne}{3}] emission for V4046 Sgr (purple) and RX J1615.3-3255 (cyan). The diagonal dotted gray line indicates equal ratios.}
    \label{fig:IIItoII}
\end{figure*}

In Figure \ref{fig:IIItoII}, we compare the predictions of our T Cha models with these data.
The equilibrium ionization should result from balancing the rate of photoionizations \citep[collisional ionization is largely negligible at wind temperatures,][]{Glassgold_2007} - which depends on the spectrum - and the rate of recombinations or charge exchange to lower ionization states - which depends largely on density.
Therefore, each panel shows a different density normalization (expressed in terms of the observable mass-loss rate), while each series of points shows a different spectrum.

We first consider low mass-loss rates $\dot{M} \lesssim 10^{-9}~M_\odot\,\mathrm{yr}^{-1}$ (upper panels of Figure \ref{fig:IIItoII}).
In the photoevaporative scenario, these are the rates typically achievable by EUV-driven photoevaporation \citep{Hollenbach_1994,Shu_1993,Wang_2017}, including considering the upper limit on the EUV photon flux of T Cha (see Section \ref{sec:methods_Mdot}). This because for EUV to drive the wind launching throughout the wind, the whole wind must be optically thin to the EUV photons, and as such must have a low density. In turn, at these low densities, only the EUV in the spectrum is effectively absorbed, while the X-rays in the spectrum will pass through the wind almost unnoticed. These EUV winds always have ionization fractions of essentially unity, and temperatures of $\sim 10^4\,\mathrm{K}$.

We see that for all spectra considered, these low mass-loss rate models have $L_{\rm [Ne\,III]}/L_{\rm [Ne\,II]}>1$ and $L_{\rm [Ar\,III]}/L_{\rm [Ar\,II]}>1$, in complete contrast to all of the data. The reason can be understood by comparing the spectra to the outer shell ionization energies of \ion{Ne}{1} ($21.56\,\mathrm{eV}$), \ion{Ne}{2} ($40.96\,\mathrm{eV}$), \ion{Ar}{1} ($15.76\,\mathrm{eV}$) and \ion{Ar}{2} ($27.63\,\mathrm{eV}$) which all correspond to photon energies in the EUV.
Since the wind has temperatures far exceeding the $\sim 1000\,\mathrm{K}$ excitation temperatures of the mid-IR emission lines, the line ratio will be proportional to the ratio of the abundances of each ion.
Moreover, the ionization balance will be between the direct photoionization by hard EUV photons of \ion{Ne}{2} to \ion{Ne}{3} (or \ion{Ar}{2} to \ion{Ar}{3}) at rate $\Phi_{23}$ and the recombination of free electrons with the ions at electron-density dependent rate $R^{\rm rec}_{32}(n_{\rm e})$:
\begin{equation}
    \frac{n_{\rm III}}{n_{\rm II}} = \frac{\Phi_{23}}{R^{\rm rec}_{32}(n_{\rm e})}
    \label{eq:ionRatioEUV}
    .
\end{equation}
Since in these highly ionized environments the electron fraction (and temperature) will have more or less saturated, the ratio of the ion abundances becomes simply a measure of the ionizing photons.
To produce abundant \ion{Ne}{2} in the wind but not \ion{Ar}{3}, one would require a spectrum with abundant photons above $21.56\,\mathrm{eV}$ but negligible photons above $27.63\,\mathrm{eV}$. Such a spectrum can only be achieved if the emitting plasma is all $\lesssim 10^4\,\mathrm{K}$, but not in the presence of hot $\sim 10^7\,\mathrm{K}$ X-ray emitting plasma - as is required to explain T Cha's X-ray spectrum - since that has a continuum tail that extends into the EUV. While one might be able to screen out the EUV by some attenuating material near to the star, the only radiation available to drive the wind would be X-rays which are expected to produce much denser winds and hence higher mass-loss rates (despite the slightly lower temperatures); such a model would not be self-consistent.

Moreover, note how, in line with Equation \ref{eq:ionRatioEUV}, the ratios predicted by the loLX\_TCha, Ercolano\_RSCVn and hiLX\_TCha spectra are so arranged in order of their EUV photon fluxes.
However, the additional $T_3=10^4\,\mathrm{K}$ EUV component added to the loLX\_TCha\_sUV and hiLX\_TCha\_sUV spectra mostly contributes flux below $20\,\mathrm{eV}$ (i.e. it is quite soft). Hence, in line with the ionization energies discussed above, it does not contribute to $\Phi_{23}$ and thus cannot contribute to the secondary ionization of Ne or Ar \citep[this is similar to the argument of][that to produce $L_{\rm [Ne\,III]}>L_{\rm [Ne\,II]}$ requires a hard EUV spectrum]{Hollenbach_2009b}. Consequently, these spectra do not produce significantly different results for the ionization in this regime from their counterparts with no explicit EUV component.

Moving to the higher densities, we see that the line ratios are much lower as the denser gas can lower its ionization through recombination or charge exchange more easily.
Henceforth we should therefore consider only winds with nominal mass-loss rates $\gtrsim 0.3 \times 10^{-8}~M_\odot\,\mathrm{yr}^{-1}$ - which will be X-ray--heated winds (though their innermost regions may be EUV--heated) - as the only self-consistent scenario able to produce line ratios that are remotely consistent with the range $0.045<L_{\rm [Ne\,III]}/L_{\rm [Ne\,II]}<0.13$ observed for most sources, including T Cha.
The highest density $\dot{M}=1-3 \times 10^{-8}~M_\odot\,\mathrm{yr^{-1}}$ models are particularly favored as they most closely cluster around the observed range.

Nevertheless these models still somewhat overpredict the Ar line ratio for T Cha. At these densities, the abundance of \ion{Ar}{3} is controlled by the balance of its production via inner (L) shell ionization by photons with energies $\gtrsim 250\,\mathrm{eV}$ and charge exchange with hydrogen (or recombination with free-electrons) forming \ion{Ar}{2}, while that of the \ion{Ar}{2} is results from the balance of its production from charge exchange between \ion{Ar}{3} and its recombination with free electrons.
Therefore, the overall ionization balance of Ar may, similar to that of Ne \citep{Glassgold_2007}, obey $\frac{n_{\rm III}}{n_{\rm II}} \propto L_{\rm X}^{1/2}$. To lower the degree of ionization, the spectrum must contain somewhat fewer soft X-ray photons, though not so few that we no longer reproduce the observed  \ion{Ar}{2} luminosity.
The weaker dependence of the \ion{Ar}{2} abundance on the X-ray luminosity means that this is a reasonable way to lower the [\ion{Ar}{3}] flux without strongly affecting the [\ion{Ar}{2}] flux.
Thus, it is likely that the spectra we utilize contain slightly too many soft X-ray photons (while having the appropriate number of hard X-ray photons $\gtrsim 870\,\mathrm{eV}$ to produce the right ratio of the Ne lines).
This is not surprising since the soft end of the spectrum is the most easily absorbed and thus the hardest to constrain accurately observationally; it could also be more easily screened between the star and disk. Since the [\ion{Ar}{3}] line is only fairly weakly detected and does not play a critical role in the arguments we make henceforth, we do not try to adjust the spectrum. 

Note that the addition of a soft EUV component at these higher mass-loss rates \textit{lowers} the line ratios (as is most easily seen by comparing for the loLX\_TCha spectra).
This results since these winds are in a partially-ionized regime, where additional soft EUV flux raises the electron density \citep[which scales with the square root of the ionization rate $\Phi$:][]{Glassgold_1997,Igea_1999,Glassgold_2007}, and thus the recombination rates of \ion{Ne}{3} to \ion{Ne}{2} and \ion{Ar}{3} to \ion{Ar}{2} \citep[while being unable to directly produce the doubly-ionized species as discussed above, see also][]{Hollenbach_2009b}. Given the low observed line ratios, it is therefore generally favourable to include the additional soft EUV component. This, along with the suggestion that our models maybe contain excessive soft X-ray, is in line with the argument presented by \citet{Bajaj_2024} that the high $L_{\rm [Ar\,II]}/L_{\rm [Ar\,III]}$ suggests EUV ionization of Ar, not (just) soft X-ray ionization.

\subsection{Line luminosities}
\label{sec:fluxes}
\begin{figure*}
    \centering
    \includegraphics[width=\linewidth]{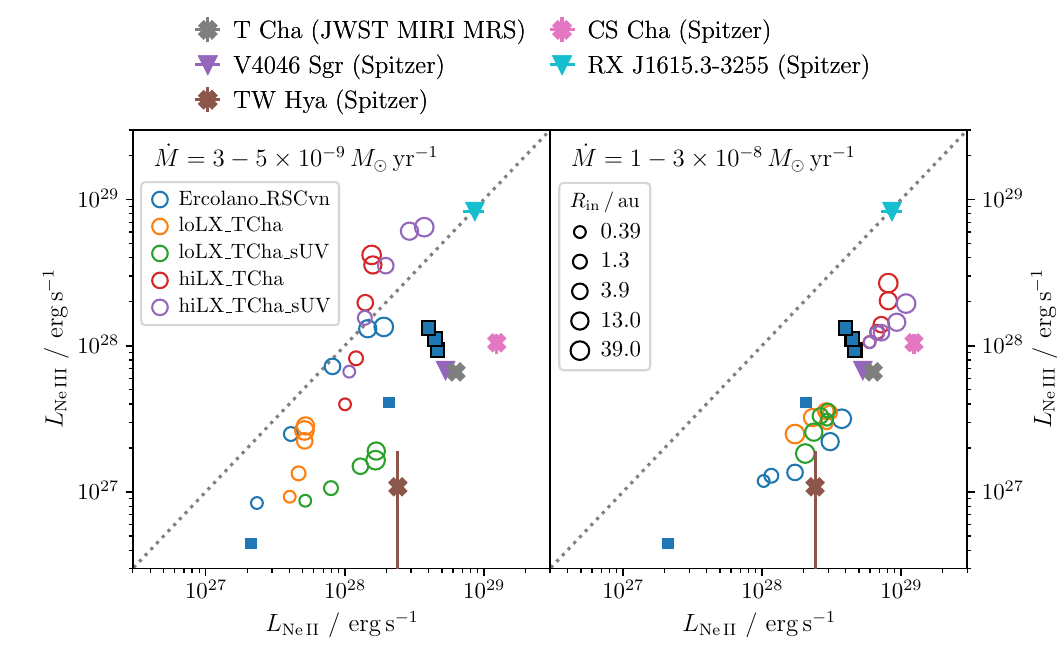}
    \caption{Comparison of the fluxes of the [\ion{Ne}{3}] $15.55~\mu\mathrm{m}$ and [\ion{Ne}{2}] $12.81~\mu\mathrm{m}$ lines. The models are plotted as for Figure \ref{fig:IIItoII}, focussing on higher density normalizations (mass-loss rates). The JWST measurements for T Cha are included for comparison (gray cross) along with literature estimates for other sources (Table \ref{tab:JWSTTCha}). The blue squares are predictions from \citet{Ercolano_2010} (borderless: full disk, black border: gas cavity) using the same spectrum as our Ercolano\_RSCVn models.}
    \label{fig:NeRatio_compare}
\end{figure*}

We can now turn our attention to the absolute luminosities, focusing, as argued above, on higher (nominal) mass-loss rates as appropriate to an X-ray regime. The luminosities of the two Ne lines are compared in Figure \ref{fig:NeRatio_compare}; for each spectrum we show models with five different inner radii (as indicated by the point size).

At each point in the wind, the line emissivities should scale with the abundance of the relevant species.
In turn, the total amount of these species across the whole wind will correlate positively with the amount of ionizing X-ray absorbed: $L_{\rm X, abs} = L_{\rm X} (1-e^{-\tau})$.
So long as the gas remains optically thin to these X-rays (as is the case for Ne as argued above), then $L_{\rm X, abs} \approx L_{\rm X} \tau \propto N L_{\rm X} \propto L_{\rm X} \dot{M}$. We therefore expect that models with both higher luminosity and high mass-loss rates should produce brighter [\ion{Ne}{2}] emission. 

Comparing the panels of Figure \ref{fig:NeRatio_compare}, we see that indeed the $1-3 \times 10^{-8}~M_\odot\,\mathrm{yr^{-1}}$ models are more luminous in [\ion{Ne}{2}] than the $3-5 \times 10^{-9}~M_\odot\,\mathrm{yr^{-1}}$ models as there is a greater column of gas to absorb the X-rays and produce ionized Ne emission. While the most luminous of the latter set do approach the [\ion{Ne}{2}] luminosity of T Cha, they do so at much too high a [\ion{Ne}{3}] luminosity as these are disks with large inner radii and low peak densities which can reach a higher degree of ionization\footnote{This is consistent with Figure \ref{fig:emission_maps} where the ionization front can be seen as a conical structure indicating that the upper latitudes of the wind become optically thin to the EUV.}.
Thus, we favour $1-3 \times 10^{-8}~M_\odot\,\mathrm{yr^{-1}}$ as the only models capable of reaching the observed [\ion{Ne}{2}] luminosity of T Cha at the right [\ion{Ne}{3}] luminosity.

Comparing the performance of the different spectra, we can see that the observed value is straddled by the higher luminosity spectra  hiLX\_TCha and hiLX\_TCha\_sUV and the other, lower luminosity, spectra.
This means that a key component of a favoured model is a high $L_{\rm X}$ as found in the hiLX\_TCha or hiLX\_TCha\_sUV spectra.

Since all of our models remain optically thin to the hard X-rays that ionize Ne, each point in the wind receives essentially the same X-ray luminosity regardless of the assumed $r_{\rm in}$ and thus total column density. Conversely, the EUV ionization is confined to a thin Str\"{o}mgren layer close to wherever most of the column density is accumulated (in our case the inner boundary of the simulation).
The EUV-ionized layer can become considerably ionized, with electron fractions close to unity i.e. $n_{\rm e} \sim n_{\rm gas}$. Given equation \ref{eq:nG} and the $r^{-1.5}$ profile in our models, the gas density at the base reaches $n_{\rm gas} \sim n_{\rm crit}$ for $r\sim 1.9\,r_G$. Thus, so long as $r_{\rm in} \lesssim 1.9\,r_G$, then in the EUV-heated region $n_{\rm e} \sim n_{\rm gas} > n_{\rm crit}$. On the other hand, the X-ray heated region has $n_{\rm e} \ll n_{\rm H}$ and so achieves $n_{\rm e} > n_{\rm crit}$ only at a radius $r_{\rm crit} \ll r_G$. Consequently, the critical density of electrons is usually reached in the EUV-ionized inner parts of the wind only and thus $r_{\rm crit} \sim r_{\rm in}$. 

The result is that the [\ion{Ne}{2}] emissivity in the X-ray--heated regions is not sensitive to critical density effects or the inner radius - the contribution of these regions decreases slightly as the inner radius is increased simply because of the loss of flux from small radii.
The much more important dependence on the inner radius comes through the EUV-heated region. Since this region is mostly super-critical, then its flux depends as $r_{\rm crit}^{3-b}$. The contribution from this thin inner ring - which can be seen clearly in the bottom row of Figure \ref{fig:emission_maps} - thus grows strongly as the inner radius is increased, only saturating at $r\gtrsim r_G$ once the gas density too low for even a fully-ionized gas to reach the critical density. 
As a consequence, increasing $r_{\rm in}$ typically has a positive effect on the luminosity at at $1-3 \times 10^{-8}~M_\odot\,\mathrm{yr^{-1}}$ (a stronger effect is seen for $3-5 \times 10^{-9}~M_\odot\,\mathrm{yr^{-1}}$ where the contribution of the EUV-heated region is relatively larger since less of the X-ray is absorbed).
However, despite this dependence, we cannot use Ne line luminosities alone to constrain $r_{\rm in}$ as whether a large or small radius is preferred depends on whether the wind spectrum has a low or high $L_{\rm X}$ respectively, and there is enough observational uncertainty in this value.

Although not tuned for either of the other systems, this set of high $\dot{M}$ models is also broadly consistent with the [\ion{Ne}{2}] and [\ion{Ne}{3}] fluxes for V4046 Sgr, CS Cha and TW Hya.
Taken at face value, TW Hya would appear to require a lower luminosity spectrum, a reasonable outcome given its later spectral type than the other stars. Moreover it is also known to have a softer spectrum in which the hardest $2-3\times10^7\,\mathrm{K}$ component usually present in the X-ray spectrum of T Tauri stars \citep[e.g.][]{Preibisch_2005} is absent \citep{Nomura_2007}. Hence it will likely produce fewer photons at $\sim 1\,\mathrm{keV}$ that are able to ionize Ne and should be studied more in future with more tailored spectra.

\begin{figure*}
    \centering
    \includegraphics[width=\linewidth]{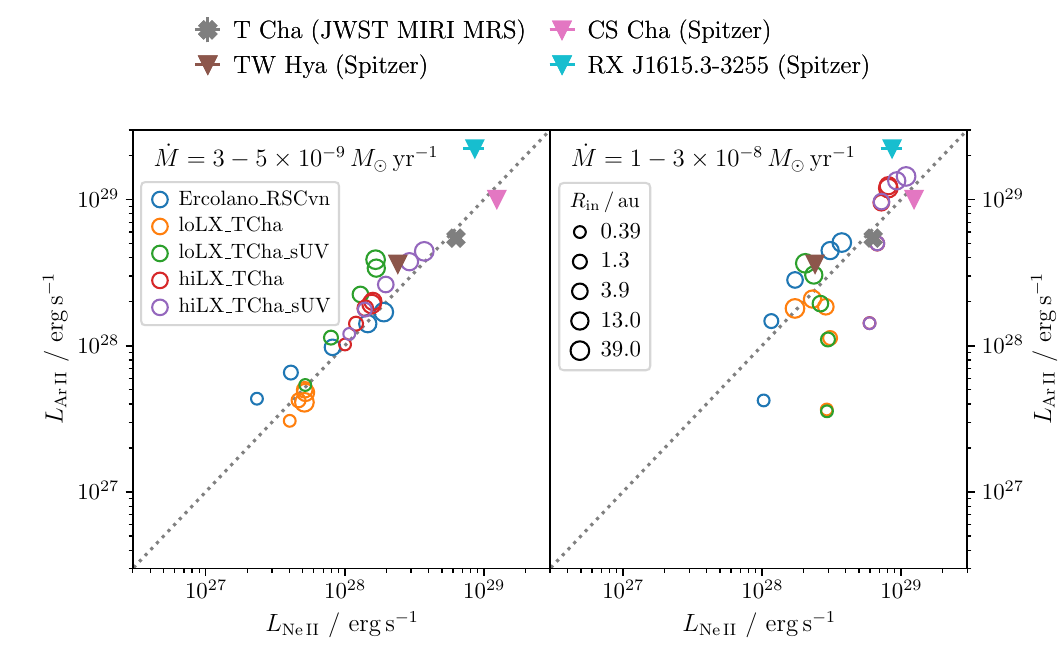}
    \caption{As Figure \ref{fig:NeRatio_compare} but comparing the ratio of the [\ion{Ar}{2}] $6.98~\mu\mathrm{m}$ and [\ion{Ne}{2}] $12.81~\mu\mathrm{m}$ lines.}
    \label{fig:ArRatio_compare}
\end{figure*}

We can get a better constraint by comparing the [\ion{Ne}{2}] and [\ion{Ar}{2}] luminosities (Figure \ref{fig:ArRatio_compare}). At $\dot{M}=1-3 \times 10^{-8}~M_\odot\,\mathrm{yr^{-1}}$, we see a much stronger dependence on the inner radius for the [\ion{Ar}{2}] than for the [\ion{Ne}{2}]. There is generally little difference between the models with and without soft EUV, showing that [\ion{Ar}{2}] is mostly produced by soft X-rays. The $r_{\rm in}$ dependence then arises since these soft X-rays are more easily absorbed, and thus the contribution of the X-ray--ionized region now depends on the assumed inner radius because of its effect on the attenuating column density (Equation \ref{eq:N_values}).
Thus, for our high $L_{\rm X}$ spectra, we find that the [\ion{Ar}{2}] flux taken in combination with the [\ion{Ne}{2}] suggests an inner radius of around $0.1\,r_G$ gives the appropriate column density. Lower luminosity spectra with large inner radii would tend to overpredict [\ion{Ar}{2}] relative to [\ion{Ne}{2}].
However one caveat is the possibility, as discussed above, that our spectra contain slightly too many soft X-ray photons, which might make the larger inner radii less unfavourable.

\subsection{Comparisons to other predictions for line ratios}
For comparison, we also include in Figure \ref{fig:NeRatio_compare} the luminosities calculated by \citet{Ercolano_2010} by postprocessing radiation-hydrodynamic simulations of both full (``primordial'') disks and those with a transparent gas cavity (``inner-hole'', which show significant emission near the midplane from the flow emanating from the directly irradiated cavity rim).
Although designated a transition disk (on account of an inner hole in its dust emission), T Cha shows no evidence for a significant inner gas cavity \citep{Wolfer_2023} and hence the appropriate model for comparison is the $L_{\rm X}=2\times10^{30}\,\mathrm{erg~s^{-1}}$ primordial disk, with $L_{\rm [Ne\,II]}=2.07\times10^{28}\,\mathrm{erg~s^{-1}}$ and $L_{\rm [Ne\,III]}=4.06\times10^{27}\,\mathrm{erg~s^{-1}}$ (indicated by the blue, borderless, square near the center of each panel). The underlying simulation had a mass-loss rate of $1.36\times10^{-8}\,M_{\sun}\,\mathrm{yr^{-1}}$.
We see that for comparable estimated observable mass-loss rates $\gtrsim 10^{-8}\,M_{\sun}\,\mathrm{yr^{-1}}$ (right-hand panel), this point lies close to our Ercolano\_RSCVn models (blue circles) with a relatively large wind radius (the closest model predictions are for an inner radius $\sim r_G$).
Thus despite the simplified hydrodynamic model we use, our results are in concordance with more sophisticated treatments of X-ray--driven photoevaporation, and suggest that despite mass-loss extending somewhat further in, the column density may be mostly accumulated around $r_G$.

\cite{Hollenbach_2009b} studied the same line ratios as here, showing how they depend on the ionizing spectrum. They argued that a ``hard-EUV' spectrum - where there are a substantial number of photons at energies above $41\,\mathrm{eV}$ (the second ionization energy of Ne) - was necessary to get $L_{\rm [Ne\,III]}>L_{\rm [Ne\,II]}$. On the contrary, a ``soft-EUV'' spectrum - such as produced by blackbody emission up to a $\mathrm{few}\times 10^4\,\mathrm{K}$ - or an X-ray spectrum - which produces weakly ionized gas - would result in $L_{\rm [Ne\,III]}<L_{\rm [Ne\,II]}$. This has been used by \citet{Szulagyi_2012} and \citet{Bajaj_2024} to argue against a hard EUV model. Moreover, due to a cancellation of the effects of many atomic parameters, $L_{\rm [Ne\,II]}/L_{\rm [Ar\,II]} \sim 1$ in either a soft-EUV case or soft--X-ray case, whereas a hard--X-ray spectrum is much better at ionizing Ne than Ar and should produce a ratio $\sim 2.5$ \citep{Hollenbach_2009b,Szulagyi_2012}.

Our spectra contain both soft and hard X-rays, as well as hard EUV (in the low-energy tail of the X-ray spectrum), and, in the case of the Ercolano\_RSCVn, hiLX\_TCha\_sUV and loLX\_TCha spectra, soft EUV from plasma at $\sim 10^4\mathrm{K}$. The outcome is thus determined by which parts of these spectra are absorbed and where. At low mass-loss rates, the wind is optically thin to the EUV and X-ray is barely absorbed so we find  ratios that are as expected for a hard-EUV model. Our higher mass-loss rate models are not well-penetrated by the EUV so most of the wind ends up with the ratios predicted by an X-ray model. When the inner radius is large, the wind is optically thin to both soft and hard X-rays, so we get $L_{\rm [Ne\,II]}/L_{\rm [Ar\,II]} \sim 1$. However Figure \ref{fig:ArRatio_compare} shows that when the inner radius becomes small enough ($r_{\rm in} \lesssim 0.1 \,r_G$), $L_{\rm [Ne\,II]}/L_{\rm [Ar\,II]} > 1$. This happens because the soft--X-rays (and EUV) are absorbed close to the star meaning that most of the wind is now absorbing only hard X-rays, thus raising the value of $L_{\rm [Ne\,II]}/L_{\rm [Ar\,II]}$, Nevertheless, in the $r_{\rm in} = 0.1 \,r_G$ case, this ratio is only just more than 1, so is still just within the range of soft X-ray models. Thus our results are all in line with those of \citet{Hollenbach_2009b}.

In conclusion, the information obtained from the line ratios should be understood as tracing which parts of the spectrum \textit{are being absorbed} and hence as a constraint on the wind's column density. Being optically thin to hard X-rays but optically thick to soft X-rays (and EUV) suggests a column density $3\times10^{20} \lesssim N \, / \,\mathrm{cm^{-2}} \lesssim 10^{22}$.

\section{Synthetic Imaging}
\label{sec:images}
\subsection{Analysis of synthetic images}
\begin{figure*}
    \centering
    \includegraphics[width=\linewidth]{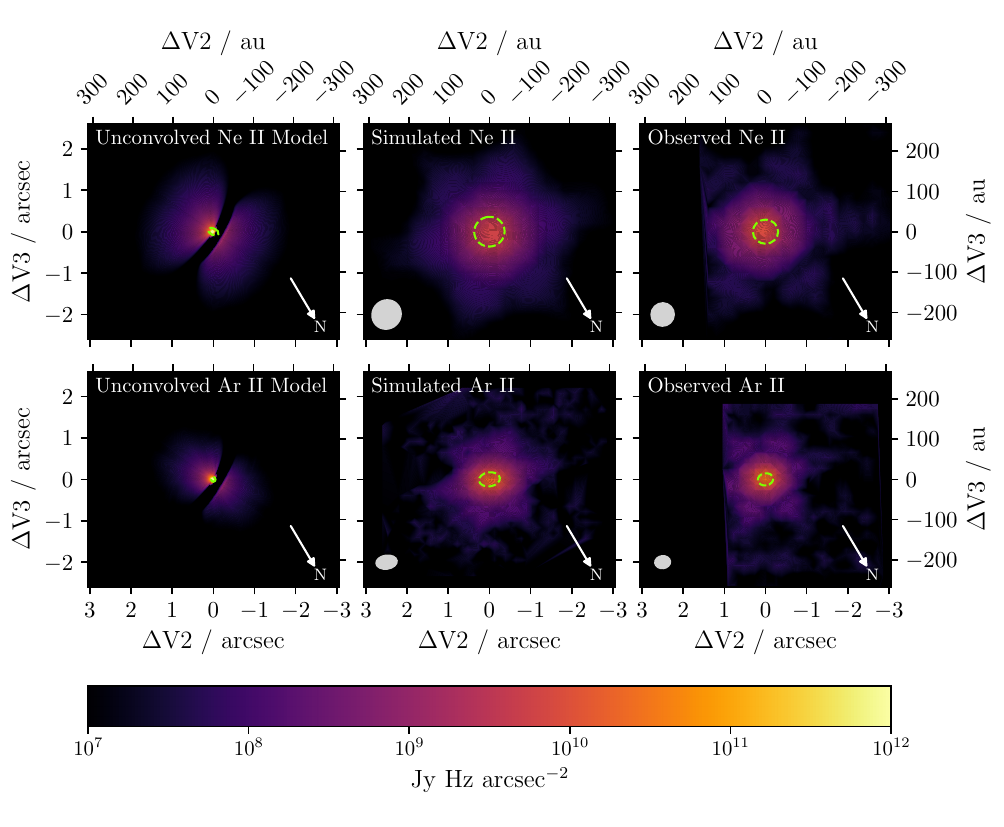}
    \caption{A comparison between the moment 0 maps of the integrated line flux across all wavelengths for the unconvolved model for the line emission (left), the continuum-subtracted synthetic image of each line (center) and the observed line emission \citet[][right]{Bajaj_2024}. Superposed as green dashed lines on the central and right-hand panels are the 2-D Gaussian fits, while the left-hand panels instead show the estimated deconvolved fit. The gray ellipses on the central and right-hand panels indicate the size of the synthetic PSF and HD37962 PSF model respectively.}
    \label{fig:syntheticImages}
\end{figure*}

We synthesize and analyze MIRI-MRS observations of the [\ion{Ne}{2}] and [\ion{Ar}{2}] lines based on our model set hiLX\_TCha\_sUV with $\dot{M}=1-3\times10^{-8}\,M_\odot\,\mathrm{yr^{-1}}$ using \textsc{MIRISim} \citep{Klaassen_2021} as described in Section \ref{sec:methods_synthetic}. We choose this set as it generally best reproduces the Ne line luminosities and thus requires minimal rescaling of the fluxes (to ensure we are equally as sensitive as the observations). We investigate models both with and without a transparent $15\,\mathrm{au}$ cavity in the dust disk (which determines whether emission from the backside can be seen or not).
Example synthetic images of the two lines for the model with $r_{\rm in}=0.1\,r_G$ and including a cavity are shown in the central panels of Figure \ref{fig:syntheticImages}. These may be compared to the unconvolved models of the lines in the left-hand panels; evidently the morphology of the underlying emission is not apparent and the image more closely resembles the PSF. The images look broadly similar to the observations presented by \citet{Bajaj_2024} of the lines for T Cha (right-hand panels), although some details of the PSF shape are somewhat different and the synthetic PSF (Gaussian fit indicated in green) is slightly larger, a known issue with \textsc{MIRISim}. Nevertheless, we will now show that useful information can still be derived from these images.

For each synthetic image, 2-D Gaussian fitting is performed using the \texttt{imfit} task from the Common Astronomy Software Application \citep[CASA,][]{McMullin_2007} package in the wavelength channels with peak line emission as described in Section 3.2.1 of \citet{Bajaj_2024}. We use a Gaussian for simplicity, to be consistent with the observational analysis in \citet{Bajaj_2024}, and to make estimates of the deconvolved size tractable. Moreover, from an observational perspective, a more appropriate morphology would not be known a priori, although since the PSF dominates the morphology, a rounded shape that can describe its core is more suitable than any function more closely describing the underlying wind morphology or PSF side lobes.
The same fitting procedure is applied to the simulated PSF at the line wavelength. Each fit is described by the Full Width at Half Maximum (FWHM) along the major and minor axes and the position angle (PA) of the major axis, which we report in Table \ref{tab:syntheticfits}. These Gaussian fits for the synthetic images and observations are superposed on the central and right-hand panels of Figure \ref{fig:syntheticImages} as the white dashed lines and by eye do an adequate job of describing the core of the emission.

Within the formula for a 2-D Gaussian, the FWHM and PA may be summarized in the correlation matrix $\mathbf{C}$. Under the assumption that the underlying emission and PSF can both reasonably be modeled by such a Gaussian, the correlation matrix of the observed (convolved) emission is simply the sum of the correlation matrices of these two (unconvolved) components. Thus, the deconvolved correlation matrix of the underlying emission may be estimated as the difference between the correlation matrices of the line emission and the PSF:
\begin{equation}
    \mathbf{C}_{\rm dec}=\mathbf{C}_{\rm obs}-\mathbf{C}_{\rm PSF},
    \label{eq:deconvolve}
\end{equation}
where each matrix has the form:
\begin{equation}
    \mathbf{C}_{i} = 
    \begin{bmatrix}
        F_{i, 1}^2 c^2_i + F_{i, 2}^2 s^2_i & (F_{i, 1}^2 - F_{i, 2}^2) c_i s_i \\
        (F_{i, 1}^2 - F_{i, 2}^2) c_i s_i & F_{i, 1}^2 s^2_i + F_{i, 2}^2 c^2_i
    \end{bmatrix}
\end{equation}
for major and minor axis FWHM $F_1$ and $F_2$ and where we use the shorthand $c_i=\cos(PA_{i})$, $s_i=\sin(PA_{i})$.

The major- and minor-axis extents of this component can be calculated as the square roots of the eigenvalues of the estimated deconvolved correlation matrix $\mathbf{C}_{\rm dec}$.
The position angle of the emission is obtained from the direction of the eigenvector of $\mathbf{C}_{\rm dec}$ corresponding to the larger eigenvalue.
Assuming the eigenvalues are both positive, then since the emitting area at half maximum of the Gaussian is given by $\frac{\pi}{4} FWHM_{\rm maj}*FWHM_{\rm min}$, we may calculate the effective emitting size from the eigenvalues as $HWHM_{\rm eff}=0.5\sqrt{FWHM_{\rm maj}*FWHM_{\rm min}}$.
If however, one of the eigenvalues is negative, our deconvolved matrix no longer represents a Gaussian. We take this as an indication that the emission is unresolved; where we require a value to include in the figures, we use an upper limit based solely on the single positive eigenvalue.

The parameters of all of the 2-D fits are reported in Table \ref{tab:syntheticfits} as the ``convolved'' values, while the values that we estimate as described in the preceding paragraphs are reported as the ``deconvolved'' values.
We also report the parameters for a fit to the continuum emission; this allows us to quote the estimated deconvolved position angle of the line emission relative to the continuum in order to account for differences between \textsc{MIRISim}'s assumed direction of North and that of the observations.

Note that we also tested using the horizontal and vertical cuts as shown in Figure 3 of \citet{Bajaj_2024} but these performed much less well at matching the observed PA, which we attribute to differences in the shape of the PSF between the \textsc{MIRISim} model and observations. The 2-D Gaussian fits are able to average over any such differences and are also somewhat more sensitive to the extended emission.

\begin{table*}[t]
        \centering
        \caption{The FHWM and PA resulting from 2-D Gaussian fitting to the (synthetic) observations of the line emission and PSF (convolved values) and that estimated for the underlying emission using the eigenvector analysis on Equation \ref{eq:deconvolve} described above (deconvolved values). For the observations, two sets of deconvolved values are given corresponding to each standard star model. To provide a reference direction, the PA of the continuum in the same wavelength channel as the [\ion{Ne}{2}] line is given for both the observations and synthetic images. $r_{\rm in}$ is the inner edge of the wind while $R_{\rm cav}$ is the size of the transparent dust cavity. The bold rows highlight the relevant observed values and the overall best-fitting model for comparison.}
        \label{tab:syntheticfits}
        \setlength\tabcolsep{1pt}
        \begin{tabular}{rcccccccc}
        \tableline
            &\multicolumn{4}{c}{[\ion{Ne}{2}]}&\multicolumn{4}{c}{[\ion{Ar}{2}]} \\
        \cmidrule(rl){2-5} \cmidrule(rl){6-9}
            &\multicolumn{2}{c}{Convolved}&\multicolumn{2}{c}{Deconvolved}&\multicolumn{2}{c}{Convolved}&\multicolumn{2}{c}{Deconvolved} \\
        \cmidrule(rl){2-3} \cmidrule(rl){4-5} \cmidrule(rl){6-7} \cmidrule(rl){8-9}
            & Major/Minor Axis & PA & Major/Minor Axis & PA & Major/Minor Axis & PA & Major/Minor Axis & PA \\
            & (mas) & ($^\circ$) & (mas) & ($^\circ$) & (mas) & ($^\circ$) & (mas) & ($^\circ$) \\
        \tableline
            HD37962  & 552/543 & $99\pm25$ & & & 366/294 & $61\pm2$ \\
            HD167060 & 557/530 & $64\pm75$ & & & 373/294 & $63\pm3$ \\
            Simulated Standard Star & 710/675 & $134\pm5$ & & & 500/320 & $98\pm1$ \\
        \tableline
            Observed Continuum & & $115\pm3$ \\
            Simulated Continuum & & $147\pm1$ \\
        \tableline
            \textbf{T Cha (HD37962)} & 614/583 & $55\pm8$ & \textbf{279/199} & $\mathbf{48}$ & 376/295 & $64\pm2$ & $\mathbf{<91}$ & $\mathbf{82}$ \\
            (HD167060)      &         &          & 263/238 & $33$ &         &          & 52/13 & $88$ \\
        \tableline
            $r_{\rm in}=0.03\,r_G$ & 767/741 & $93\pm10$ & 343/245 & $65$ & 547/356 & $98\pm1$  & 222/156 & $98$\\
            $r_{\rm in}=0.1\,r_G$  & 770/743 & $95\pm8$  & 348/254 & $66$ & 537/346 & $98\pm1$  & 196/132 & $98$ \\
            $r_{\rm in}=0.3\,r_G$  & 798/767 & $92\pm10$ & 406/317 & $68$ & 562/373 & $99\pm1$  & 257/191 & $104$ \\
            $r_{\rm in}=1\,r_G$    & 785/753 & $112\pm7$ & 361/306 & $79$ & 619/440 & $102\pm1$ & 369/297 & $115$ \\
            $r_{\rm in}=3\,r_G$    & 1139/956& $63\pm2$  & 915/644 & $61$ & 1192/866 & $70\pm2$ & 1100/780 & $64$ \\
        \tableline
            $r_{\rm in}=0.03\,r_G$, $R_{\rm cav} = 15\,\mathrm{au}$ & 738/716 & $102\pm10$ & 271/173 & $64$ & 519/345 & $97\pm1$  & 145/123 & $66$\\
            $\mathbf{r_{\rm in}=0.1\,r_G}$,  $\mathbf{R_{\rm cav} = 15\,{\rm au}}$ & 740/718 & $103\pm11$ & \textbf{267/162} & $\mathbf{64}$ & 512/340 & $98\pm1$  & \textbf{115/110} & $\mathbf{8}$ \\
            $r_{\rm in}=0.3\,r_G$,  $R_{\rm cav} = 15\,\mathrm{au}$ & 768/742 & $94\pm9$   & 345/289 & $65$ & 536/363 & $98\pm1$  & 193/171 & $98$ \\
            $r_{\rm in}=1\,r_G$,    $R_{\rm cav} = 15\,\mathrm{au}$ & 780/750 & $109\pm8$  & 354/293 & $75$ & 597/436 & $101\pm2$ & 331/291 & $120$ \\
            $r_{\rm in}=3\,r_G$,    $R_{\rm cav} = 15\,\mathrm{au}$ & 1136/953& $63\pm2$   & 911/639 & $61$ & 1180/857 & $69\pm2$ & 1088/769 & $63$ \\
        \tableline
        \end{tabular}
\end{table*}

\subsection{Resolved [\ion{Ne}{2}]}
\begin{figure*}[ht]
    \centering
    \includegraphics[width=\linewidth]{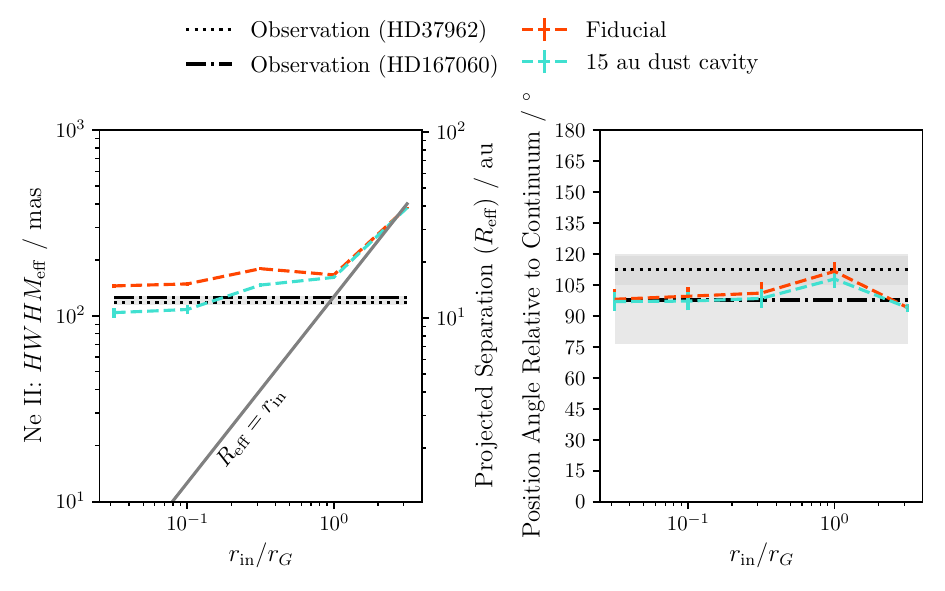}
    \caption{The estimates of the deconvolved $HWHM_{\rm eff}$ (left) and position angle relative to the continuum (right) of the [\ion{Ne}{2}] $12.81\,\mu\mathrm{m}$ emission in our models without (fiducial, red) and with a 15 au dust cavity (cyan) as a function of the model's inner radius, estimated from 2-D Gaussian fits. For the left-hand panel, the right-hand axis gives the corresponding deprojected size at the distance of T Cha. The horizontal dotted lines indicate the observational value derived using the standard star HD37962 as the PSF model, while the dot-dashed lines are the equivalent derived using HD167060 as an alternative standard star. The errorbars on the simulation data (or equivalently the shaded gray regions for the observations) are the uncertainties obtained by propagating the uncertainty on the position angle of the fit; these are typically smaller or similar in size to the systematic difference due to the uncertainty in the PSF model.}
    \label{fig:HWHM_NeII}
\end{figure*}

In Figure \ref{fig:HWHM_NeII} we show the results of the Gaussian fitting to the [\ion{Ne}{2}] $12.81\,\mu\mathrm{m}$ emission for our hiLX\_TCha\_sUV model set as a function of $r_{\rm in}$, with uncertainty estimates produced by propagating the uncertainty on the PA in each Gaussian fit.
To contextualize the values, the right-hand axis shows the calibration to a projected distance $R_{\rm eff}$ from T Cha, assuming a distance of 102.7 pc.
The horizontal black lines represent the value from the real observations using two different standard stars -  HD37962 and HD167060 - as models for the PSF for the deconvolution; comparison to the propagated statistical errors suggests that the PSF model is likely the dominant source of uncertainty in our derived values.
The estimated sizes correspond to emission on a scale of $12-13\,\mathrm{au}$, a value which lies intriguingly close to T Cha's gravitational radius.

To facilitate further interpretation, the solid gray line in each figure panel indicates where the projected separation of the effective HWHM $R_{\rm eff}$ (from now on, the ``effective radius'') is equal to the inner radius of the model.
For $r_{\rm in} \gtrsim r_G$, the $R_{\rm eff}$ of the synthetic images lies close to this line. This suggests that for winds restricted to large radii, the extent of the emission traces the inner radius of the wind: this is the case discussed in Sections \ref{sec:methods_radTrans} and \ref{sec:fluxes} where a bright inner ring of EUV-heated material near the inner radius dominates the emission.

One reason for targeting T Cha was that the continuum emission shows a large $\sim33\,\mathrm{au}$ cavity at \textit{mm} wavelengths \citep{Francis_2020}, equivalent to $\sim3\,r_G$.
Since the [\ion{Ne}{2}] profile has a net blueshift, the redshifted wind from the far side of the disk must be obscured by dust, which may imply little-to-no emission at radii inside the \textit{mid-IR} dust cavity. Thus, if the dust cavity were the same size at mid-IR wavelengths, we should expect to resolve the emission.
Our models confirm that if the wind's [\ion{Ne}{2}] emission only started beyond 30 au then we would indeed be able to recover this scale from the observations.
However the observations suggest emission that is more compact $R_{\rm eff} \lesssim r_G$ (as expected for photoevaporation from a full disk).
Moreover \citep[although not believed to be the case for T Cha,][and hence not included in our model]{Wolfer_2023} - even if there were an optically thin gas cavity in the underlying disk extending to radii $R_{\rm cav} \gtrsim 30\,\mathrm{au}$, emission could still be found at radii $R<R_{\rm cav}$ \citep[e.g.][]{Ercolano_2010} since material launched sonically from the cavity wall flows inwards initially, reaching roughly $r_G$ \citep{Alexander_2006a,Owen_2010}.

On the other hand, we see that (for cases both with or without a dust cavity) there is a very shallow dependence of $R_{\rm eff}$ on $r_{\rm in}$ for $r_{\rm in} \lesssim r_G$, with $R_{\rm eff} \sim 10-20\,\mathrm{au}$ in all cases. This means that the extent of the \ion{Ne}{2} emission can only place a relatively unconstraining upper limit on the innermost extent of the wind, and a much more compact inner radius $<r_G$ could also be consistent with the observations. That said, given the relatively small uncertainties, it may still be possible to distinguish: the closest agreement is with our model with $r_{\rm in}=0.1\,r_G$ and including a $15\,\,\mathrm{au}$ dust cavity, consistent with the rest of our picture of this source.

The reason for this shallow dependence is that, as discussed in Section \ref{sec:fluxes}, the extended emission due to X-ray photoionization dominates for small inner radii. Indeed the deconvolved Gaussian estimate overlaid on Figure \ref{fig:syntheticImages} shows that the size retrieved contains a little more than just the bright central core, indicating that extended emission contributes to the effective size.
Since it can be argued that a) for our density normalization, the strongly inner-radius--dependent EUV-ionized component stops dominating the flux for $r_{\rm in}\lesssim r_G$ on the basis of critical density arguments and b) once the extended X-ray--ionized component comes to dominate there is weak dependence on inner radius as all the models are optically thin to the $>870.1\,\mathrm{eV}$ photons needed to ionize Ne, then it is logical that the scale that is perserved is similarly on the order of $r_G$. While the X-ray--induced emission has no direct sensitivity to this scale, the intrinsic scale of this emission would have to be much smaller - and its flux contribution correspondingly lower - for the strongly $r_{\rm in}$-dependent EUV component to dominate and produce an overall $r_{\rm in}$ dependence at much smaller sizes.

The measured sizes thus suggest that the [\ion{Ne}{2}] emission traces an X-ray--ionized wind extended to at least $\sim12\,\mathrm{au}$. We can achieve this in our models with appropriate densities and inner radii such that the column of material inside $12\,\mathrm{au}$ is no more than $\sim10^{22}\,\mathrm{cm^{-2}}$; the agreement with the observed scale suggests that these models are a decent representation of T Cha and that the same constraint on column density applies also to the observations. Were the column density on these scales any higher, then Ne could only be ionized (and thus produce $12.81\,\mu\mathrm{m}$ emission) at radii smaller than those observed. This would happen in the case of a higher mass-loss rate (or the same mass-loss rate spread over a smaller extent thus resulting in a higher density) or an even smaller inner radius than those explored.

It is notable that all the $r_{\rm in}$ that can be consistent with the data are $\lesssim r_G$ and thus also $\lesssim R_{\rm cav}\approx15\,\mathrm{au}$ inferred from the SED fits. Thus, we also showed in Figure \ref{fig:HWHM_NeII} (as the cyan points) the effective radius for models with a transparent dust cavity (this is simply a condition on which parts of the wind can be \textit{seen}, we do not change the gas structure) and noted above that a model with a cavity is the best fit for the measured size.
For $r_{\rm in} \gtrsim r_G$, the results are unaffected since most or all of the emission originates outside of the cavity radius and hence cannot be seen through the cavity.
However, for smaller $r_{\rm in}$, the radius is reduced as the emission becomes more centrally concentrated in the presence of a cavity. The fact that a change can be seen emphasizes that the measured size is also sensitive to extended emission on scales larger than the cavity: if we were only measuring emission inside the cavity radius then hiding emission from the backside with an optically thick disk would not make the emission more extended but simply fainter.

In the right-hand panel of Figure \ref{fig:HWHM_NeII} we also show the PA of the emission.
The observations show a PA offset of $113^\circ$ and $98^\circ$ between the wind and disk when HD37962 and HD167060 respectively are used as the PSF model. The simulations - with or without a dust cavity - are generally in very good agreement with HD167060 while the $r_{\rm in}=0.1\,r_G$ has better agreement with HD37962.
In all cases, these values are relatively close to $90^\circ$ and the deconvolved Gaussian estimate overlaid on Figure \ref{fig:syntheticImages} is clearly approximately perpendicular to the disk plane. This is reasonable for a source viewed close to edge-on since winds involve the launch of material away from the disk plane; cancellations due to symmetry about the rotation axis should then lead to an average PA of $90^\circ$.
Note that since for any morphology which is symmetric about the rotation axis, the PA should average out to either $0^\circ$ or $90^\circ$; the fact that we retrieve a value consistent with one of these is a vindication of the 2-D Gaussian fits and our deconvolution estimate being a sufficient way to derive information about the intrinsic emission structure.
Moreover, we also consider that $90^\circ$ is the direction separating the red- and blue- shifted lobes of emission from the backside and frontside of the disk respectively. However, the two lobes are clearly not resolved in the synthetic image, and the recovered scale is slightly less than their separation. This suggests that the PA and effective size are not negatively affected by this bipolar structure, in the worst case it is simply part of the extended emission component and does not prevent a Gaussian being a sufficiently good fit to retrieve sensible information about the overall presence of this component.

However, the PA values do seem to be somewhat biased towards slightly larger angles than $90^\circ$. This suggests some asymmetry is at hand; one possible physical explanation may be that the fits are performed on the brightest wavelength channel at $12.8138\,\mu\mathrm{m}$, which is slightly redward of the rest wavelength and thus also the blueshifted line centroid.
Given the sense of the rotation of the disk \citep{Huelamo_2015}, the SW of the image is slightly less blueshifted - as this material is rotating away from us - than the SE of the image - which is rotating towards us. Though the line is spectrally unresolved, the SW should still contribute slightly more on the redshifted side of the line; the rotation of the wind material could thus be biasing the fits towards the SW of the image and increasing the PA from $90^\circ$ slightly. On the other hand, if, unlike the Gaussian assumed in the fitting, the PSF is not symmetric about the major axis, then this could also bias the measured PA.

\subsection{Unresolved [\ion{Ar}{2}]}
\begin{figure*}
    \centering
    \includegraphics[width=\linewidth]{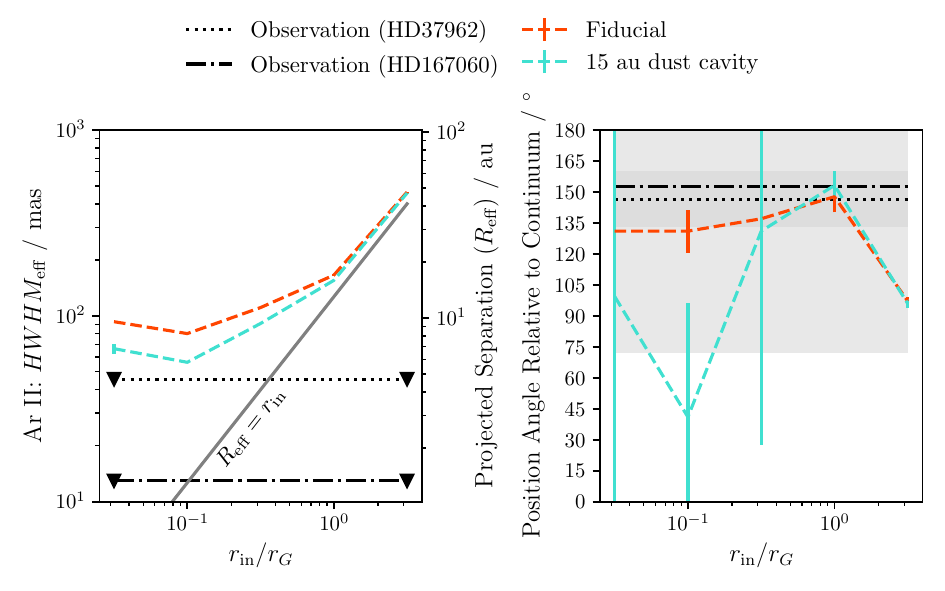}
    \caption{As with Figure \ref{fig:HWHM_NeII} but for [\ion{Ar}{2}] $6.98\,\mu\mathrm{m}$ emission. No extension is detected for either PSF: we indicate this by marking the endpoints of the line with triangles.}
    \label{fig:HWHM_ArII}
\end{figure*}

For [\ion{Ar}{2}], no extension is detected in the observations according to our deconvolution method and eigenvalue analysis for either PSF. As such, although a PA can be calculated for the eigenvector corresponding to the positive eigenvalue, it is not very physical and is very uncertain as indicated by the large shaded error regions.

For models with larger $r_{\rm in} \gtrsim r_G$, Figure \ref{fig:HWHM_ArII} shows similar behavior for [\ion{Ar}{2}] as seen for the [\ion{Ne}{2}], where the effective radius is a good measure of the inner radius. These are cases with a significant contribution from the EUV-ionized inner regions. However, while the dependence of $R_{\rm eff}$ on $r_{\rm in}$ initially flattens off a little at small radii, there is more dependence than was seen for [\ion{Ne}{2}]. The closest agreement with the observed sizes is again for $r_{\rm in} = 0.1\, r_G$, though the larger value seen for $r_{\rm in} = 0.03\, r_G$ possibly results from sensitivity issues.
The position angles have very large error bars since the emission is barely resolved allowing essentially any orientation of the underlying emission to be consistent with the synthetic image.

Earlier, we argued that the line ratio of [\ion{Ne}{2}] and [\ion{Ar}{2}] was at the higher end of what is considered achievable with soft X-rays \citep{Hollenbach_2009b} and best reproduced if we were moving towards a regime where a large part of the outer wind is only penetrated and ionized by hard X-rays and soft X-rays are screened out by its inner parts.
The behaviour seen here also occurs because unlike \ion{Ne}{2}, which is produced by hard ($\gtrsim 870 \,\mathrm{eV}$) X-rays which can penetrate all our wind models, \ion{Ar}{2} is produced by soft ($\sim 250\,\mathrm{eV}$) X-rays (and EUV). For $r_{\rm in} \leq 0.1\, r_G$, the wind becomes optically thick to these photons, which penetrate only its inner parts, thus restricting the emitting area of [\ion{Ar}{2}] significantly (and reducing the line flux).
The overlaid deconvolved Gaussian estimate in Figure \ref{fig:syntheticImages} illustrates this point that in this case the emission is dominated by the bright inner core, with any extended component being relatively insignificant.
This makes the [\ion{Ar}{2}] emission more strongly dependent on the inner radius, and thus a more sensitive probe of the inner parts of the wind where the column density is accumulated than the [\ion{Ne}{2}].
Indeed, the extent provides stronger evidence than the line fluxes as it is not subject to uncertainties in the elemental abundances, which are challenging to determine even for our own Sun \citep[see][for a discussion]{Asplund_2021}.

\begin{figure*}
    \centering
    \includegraphics[width=\linewidth]{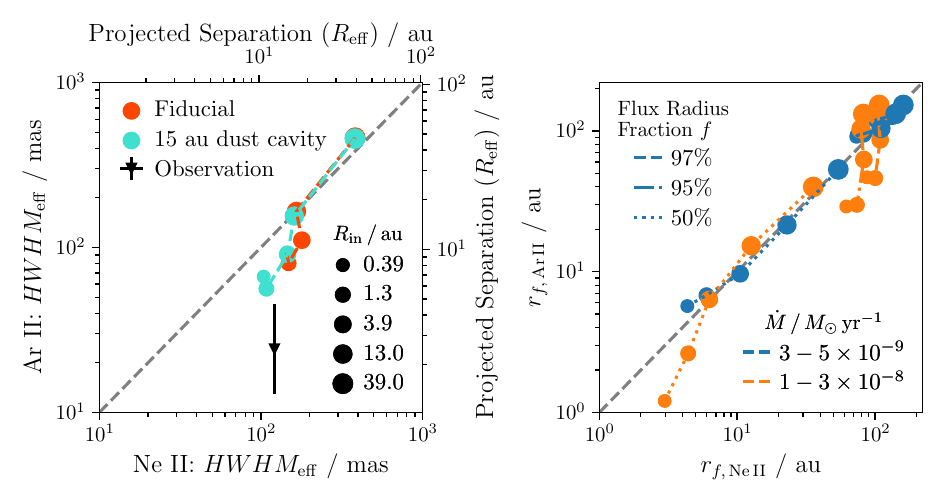}
    \caption{Left: A comparison between the estimated intrinsic $HWHM_{\rm eff}$ of the [\ion{Ne}{2}] $12.81\,\mu\mathrm{m}$ emission (x-axis) and [\ion{Ar}{2}] $6.98\,\mu\mathrm{m}$ emission (y-axis) for the fiducial set of synthetic images. The average of the values obtained from the observations with the two standard stars is indicated by the black marker with the error bars calculated as half their difference. Upper limits (no extension detected) are indicated by triangles. Right: a comparison between spherical radius containing 97 per cent (dashed), 95 per cent (dot-dashed) and 50 per cent (dotted) of the [\ion{Ar}{2}] and [\ion{Ne}{2}] flux shown for model sets with two different mass-loss rates. In each panels point sizes indicate the inner radius of the model.}
    \label{fig:HWHM_vs}
\end{figure*}

Figure \ref{fig:HWHM_vs} demonstrates that when the combination of the emitting extents of [\ion{Ne}{2}] and [\ion{Ar}{2}] are used together, the [\ion{Ar}{2}] being less extended than the [\ion{Ne}{2}] only occurs when $r_{\rm in} \lesssim 0.3\, r_G$, thus allowing [\ion{Ar}{2}] to help break the degeneracy between different $r_{\rm in}$ present when considering [\ion{Ne}{2}] alone.
Through the right-hand panel of the figure, we confirm that this is a behaviour of the underlying simulations which is preserved in the synthetic images.
We show the spherical radius enclosing different fractions $f$ of the flux in the model before it is converted into a synthetic image, including the 97 per cent and 95 per cent levels used to defined the observable mass-loss rate (Section \ref{sec:methods}). We also show the 50 per cent level, which is the flux contained within the HWHM of a 2-D Gaussian, and is thus the most direct comparison to the parameters we infer from the synthetic images. For each f, we see the same picture as for the effective radii resulting from the deconvolution: inner radii $r_{\rm in} \lesssim 0.1-0.3 \,r_G$ (and also higher mass-loss rates) are required to produce [\ion{Ar}{2}] which is intrinsically more compact than the [\ion{Ne}{2}].
This can also be seen directly in the emission maps in Figure \ref{fig:emission_maps}: only a high mass-loss rate and small inner radius (third row) produces a significant difference between the extent for the two lines. Moreover, comparing the $f=95\%/97\%$ radii to the $f=50\%$ radius reveals significant emission from scales $\sim10\times$ that traced by the effective radius.

\section{Discussion}
\label{sec:discussion}
\subsection{Consistency with line profiles}
\label{sec:profiles}
Constraints on the origin of line emission can also be derived from spectrally-resolved line profiles. In particular, the (deprojected) half width at half maximum is an indicator of the innermost emitting region under the assumption of Keplerian broadening. We thus check how well our preferred models agree with high-resolution [\ion{Ne}{2}] ($R=30000$) and [\ion{O}{1}] ($R=45000$) line profiles of T Cha \citep{Pascucci_2009,Pascucci_2020}.

\begin{figure*}
    \centering
    \includegraphics[width=\linewidth]{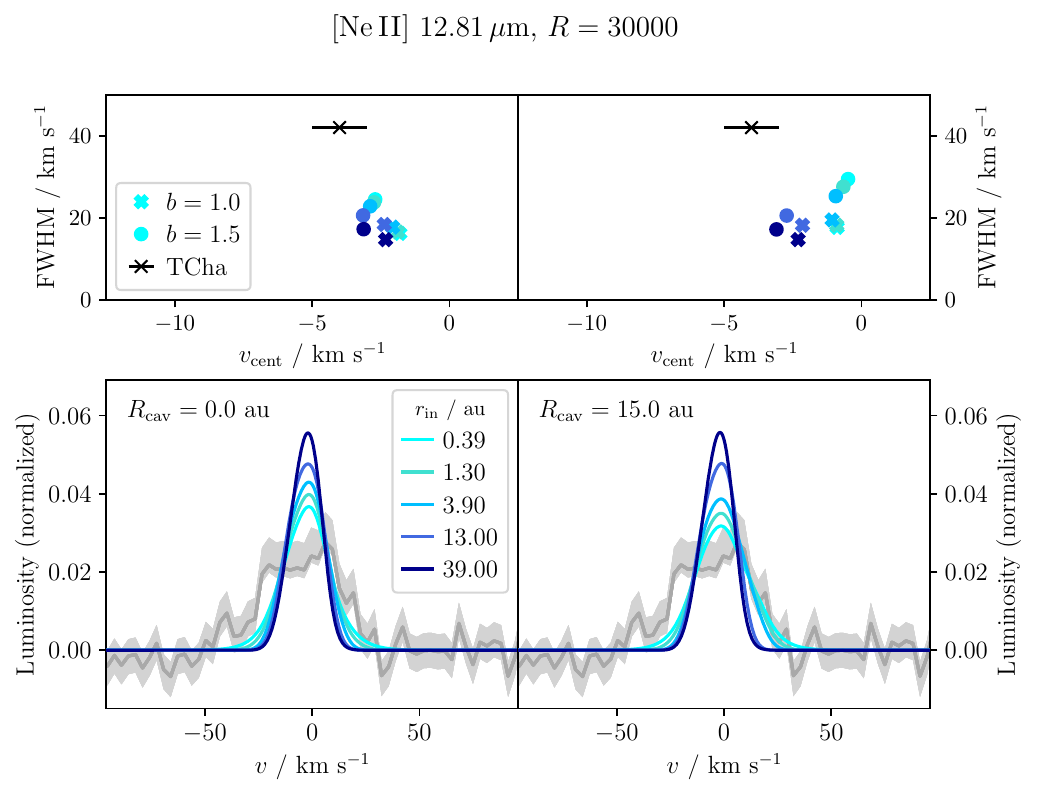}
    \caption{Bottom panels: the predicted $R=30000$ [\ion{Ne}{2}] line profiles (colored lines) without (left) and with (right) a $15\,\mathrm{au}$ dust cavity compared to the data from \citet{Pascucci_2020} (gray). Each line represents a model with a different inner radius (with increasingly dark colors indicating larger values) for a mass-loss rate of $1-3\times10^{-8}\,M_{\sun}\,\mathrm{yr^{-1}}$ and the hiLX\_TCha\_sUV spectrum. Both the models and the data are normalized to an integrated value of unity. Top panels: the summary statistics (full width at half maximum and centroid shift) of the modeled lines compared to those of the data (black cross). The colors are as in the bottom panel, with circles being for $b=1.5$ and crosses for $b=1.0$.}
    \label{fig:profile_NeII}
\end{figure*}

\begin{figure*}
    \centering
    \includegraphics[width=\linewidth]{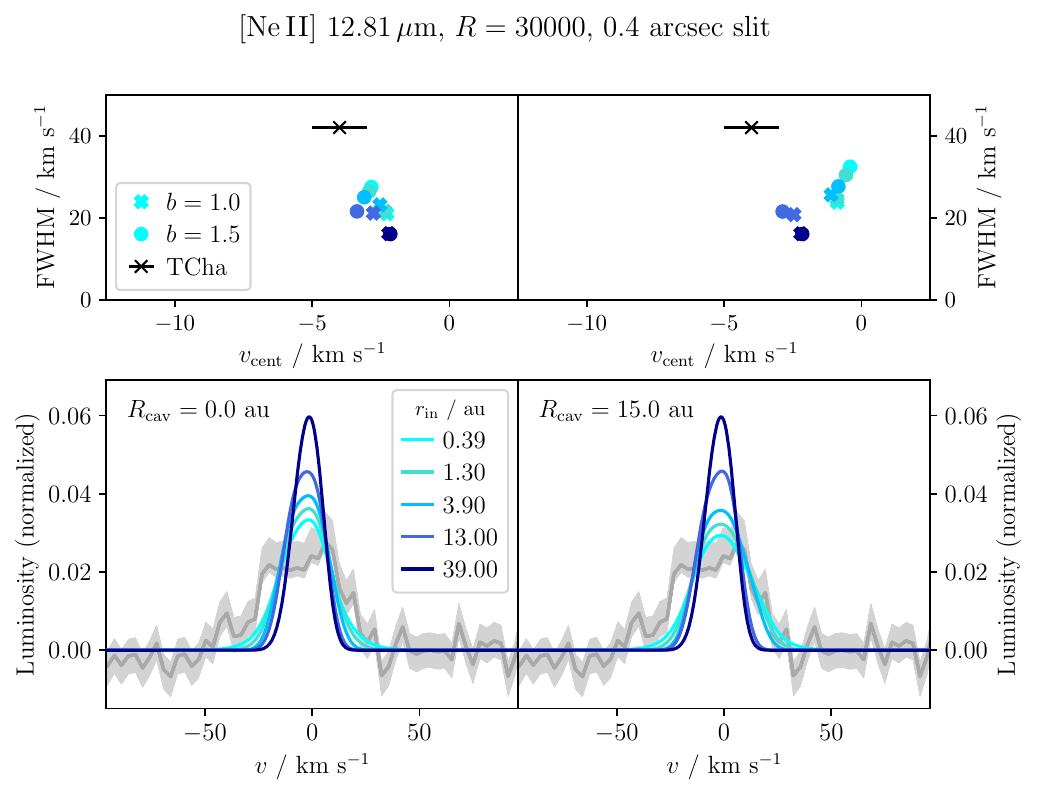}
    \caption{As with Figure \ref{fig:profile_NeII} but excluding emission from outside a 0.4 arcsec slit oriented N-S.}
    \label{fig:profile_NeII_slit}
\end{figure*}

Initially we assume that the disk is optically thick throughout the midplane and the complete emission from the near side only is seen (Figure \ref{fig:profile_NeII}, left panels). We then consider two further effects. Firstly, we assess how the presence of an optically thin dust cavity at $<15\,\mathrm{au}$ affects the line profile (Figure \ref{fig:profile_NeII}, right panels). Secondly, we reassess both of these cases taking into account the finite $0.4$ arcsec width of the slit (Figure \ref{fig:profile_NeII_slit}); since the position angle of the slit was $0^\circ$ \citep{Pascucci_2009} and that of the disk is $113^\circ$ (i.e. fairly close to perpendicular to the slit) then this means that emission separated by more than $\sim 20\,\mathrm{au}$ along the major axis may be missed.

In Figure \ref{fig:profile_NeII} we see that as one would expect, especially for such a highly inclined disk, the smaller the inner radius the broader the line profile. However, the modeled line profiles in the bottom panels when no dust cavity is included are visibly narrower than observed for all inner radii. This is confirmed in the upper panels where we compare the FWHM and centroid shift measured from our profiles to that of the data \citep[as calculated by][]{Pascucci_2020}; more encouragingly, our profiles are relatively consistent within uncertainties with the centroid blueshift. Nevertheless, the broad profile supports the case favoured in Sections \ref{sec:ratios} and Sections \ref{sec:images} where the inner radius $r_{\rm in}$ of the wind is small ($\sim 0.1 \,r_G$).

However, for the cases with $r_{\rm in}<R_{\rm cav}$, the introduction of a dust cavity leads to significant emission from the receding wind becoming visible. While this helps a little to broaden the profiles, it also shifts them towards no net centroid shift. Since the fractional error on the centroid shift \citep[which for bright lines is dominated by the uncertainty of the stellar radial velocity,][]{Pascucci_2020} is larger than that on the FWHM, this may be a worthwhile trade off. We also note that the normalized line profiles with a cavity actually agree very well with the red wing of the observed line for $r_{\rm in}\leq3.90\,\mathrm{au}$ - our poor fit is a result of too strongly peaked emission near the line center and not enough on the blue wing of the line i.e. insufficient emission at high velocities.
Although T Cha is thought to have a smaller inner disk of $\lesssim 1\,\mathrm{au}$, and even though our modeling prefers more compact inner radii - where there is emission on these scales, in practice since the emission is extended out to $\sim 10\,\mathrm{au}$, then plenty of the red-shifted emission will still be visible through the cavity and its effect on the line shape is not significantly reduced.

It is thus clear that T Cha's [\ion{Ne}{2}] line is broader than we can explain; this is consistent with previous works where it seems to be an outlier among similar disks in its breadth \citep{Pascucci_2020,Ballabio_2020}.
It is not obvious how to create more emission at higher velocities within a photoevaporative model where the line kinematics are closely tied to the sound speed. In principle, part of the answer could be a greater sound speed: \citet{Ballabio_2020} found that $c_{\rm S}=10\,\mathrm{km\,s^{-1}}$ was preferred for the [\ion{Ne}{2}] line. However such values are only achieved globally in EUV-driven winds, which we have ruled out on the basis of line ratios. 
Our sound speed is already typically $c_{\rm S} \approx 7 \,\mathrm{km\,s^{-1}}$ averaged across the wind, and our model already accounts for EUV heating raising the temperature in the inner parts and uses the temperature from the photoionization calculations to determine the sound speed locally.
Thus there is limited scope to perfect the agreement with the data with hotter gas. On the other hand, emission could be removed from the line center if for example the outer regions are cooler than assumed here and do not reach the $1123\,\mathrm{K}$ excitation temperature of the $12.81\,\mu\mathrm{m}$ line \citep{Sellek_2024b}.

Conversely, if a magnetocentrifugally launched wind achieves poloidal velocities $v_p \sim v_{\rm K}(r<r_G)$ then these will by definition be greater than the sound speed since $r_G$ is defined as the location where the Keplerian orbital velocity and sound speed are equal. While the significant amount of material inside $r_G$ suggested by our modeling probably can be explained with photoevaporation, if magnetocentrifugal forces contribute instead of or as well as thermal pressure force, the deviation of the kinematics there from those of a purely thermal wind may be sufficient to explain the observed line profile being broader and more blueshifted than our models can achieve. As simulations of MHD winds continue to be developed and improved - particularly with regards to their themochemistry - they should be tested against these sorts of observations to see if they indeed provide a remedy. \citet{Nemer_2023} provided a first step towards this using the simple analytic magnetothermal model of \citet{Bai_2016} and indeed see that such models produce much a much more extended blue wing of the [\ion{Ne}{2}] line than the same models in the purely photoevaporative limit. Nevertheless these models underpredict the luminosity of the line by an order of magnitude compared to typically observed values.

The finite slit width may also contribute to the overprediction of flux near the line center. The material excluded by the narrow slit should be mostly at large radii in the disk where the Keplerian broadening is less, meaning this emission should appear near the line center. Indeed in Figure \ref{fig:profile_NeII_slit} we see that with-or-without a dust cavity, removing this emission makes the profiles more flat-topped. It therefore also increases the FWHM of the line as a result of lowering the peak flux and making the line less centrally concentrated.

Note that in Figures \ref{fig:profile_NeII} and \ref{fig:profile_NeII_slit} we also show the FWHM and centroid shift for profiles for models with $b=1.0$ as the square points. These have a shallower density profile and consequently relatively more material (and attenuation of X-rays) at larger radii and less at smaller radii. This results in less Keplerian broadening of the line and thus a smaller FWHM, increasing the tension between our models and the data. It is for this reason that we did not further explore such models in this work.

\begin{figure*}
    \centering
    \includegraphics[width=\linewidth]{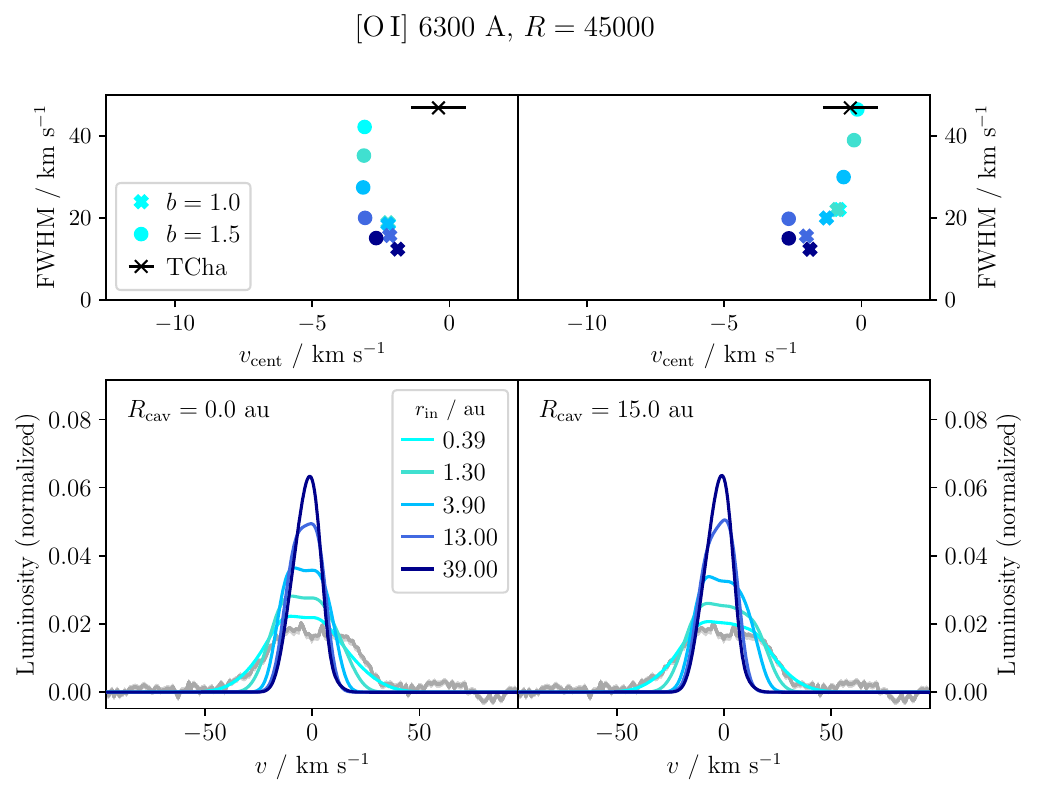}
    \caption{As with Figure \ref{fig:profile_NeII} but for the [\ion{O}{1}] line and shown at a resolution of $R=45000$ to match the data.}
    \label{fig:profile_OI}
\end{figure*}

Finally - although the focus of this work is on the mid-IR emission from Ne and Ar - in Figure \ref{fig:profile_OI} we also check the consequences of our models for the [\ion{O}{1}] $6300$ \r{A} line\footnote{Since the [\ion{O}{1}] emission is much more compact it is barely affected by the finite slit so we do not account for slit width here.}. We compare them to the $R=45000$ profile analyzed by \citet{Pascucci_2020}. While the models without a dust cavity in this case overpredict the blueshift, the presence of a cavity makes them consistent with the very low observed blueshift of the [\ion{O}{1}] line \citep[as argued for TW Hya by][]{Pascucci_2011}. Moreover, with-or-without a cavity, since in all cases the emission is constrained to the hottest inner parts of the wind, there is an even stronger dependence of the FHWM on $r_{\rm in}$. The full line profiles show that our model actually does a surprisingly good job of reproducing the observed flat-topped profile for radii $\lesssim 0.3 \,r_G$. The model with the smallest $r_{\rm in}=0.03 \,r_G$ and a dust cavity does the best job, almost exactly reproducing the line profile. $r_{\rm in}=0.1 \,r_G$ does a similarly good job of matching the shape and width of the central parts of the line, though misses flux in the line wings. The data strongly prefer $b=1.5$ over $b=1.0$. Thus, the broad [\ion{O}{1}] emission also supports the scenario of a small $r_{\rm in}$ in the disk wind from T Cha. We reiterate that T Cha is not the typical disk, and that like other transition disks with [\ion{Ne}{2}] LVCs, it lacks a [\ion{O}{1}] HVC and the [\ion{O}{1}] LVC can be fit by a single component, rather than necessitating separate broad and narrow contributions with potentially separate origins.

\subsection{What sort of wind do Ne and Ar trace in T Cha?}
Firstly we restate the conclusions of our models.
The low ratio of the line luminosities of double-ionized species to those of singly-ionized species excludes low-density winds such as might result from EUV photoevaporation.
The bright [\ion{Ne}{2}] and [\ion{Ar}{2}] lines requires a wind of reasonably high density/mass-loss rate $\gtrsim 10^{-8}\,M_{\sun}\,\mathrm{yr}^{-1}$ which is photoionized by a sufficiently large X-ray luminosity $\sim 3.7\times10^{31}\,\mathrm{erg\,s^{-1}}$.
Finally, in order to achieve $L_{\rm [Ne\,II]}/L_{\rm [Ar\,II]}$ moderately less than unity, and, crucially, more compact [\ion{Ar}{2}] emission than [\ion{Ne}{2}], then assuming a power-law wind profile with a sharp cut-off at an inner radius $r_{\rm in}$, we require a small enough inner radius $r_{\rm in} \lesssim 0.1 \,r_G$ such that photons with energies $\sim250\,\mathrm{eV}$ are screened from reaching the outer reaches of the wind by its inner parts.

Note that this large $L_{\rm X}$ is just within the range of values quoted in the literature \citep{Sacco_2014}. Additionally we note that the [\ion{Ne}{2}] emission observed with JWST is around 55 per cent brighter than the Spitzer value \citep{Bajaj_2024} which may indicate variability in the X-ray flux reaching the disk, either because the X-ray luminosity has itself increased or because the shadowing of the outer disk by an inner ``warp'' \citep{Xie_2023} has reduced\footnote{Note that an even larger change in the [\ion{Ne}{2}] and [\ion{Ne}{3}] line fluxes has been reported by \citet{Espaillat_2023} for the SZ Cha system, which they attributed to variability in an inner disk wind. They argue this has changed the dominant ionization in the outer disk wind from X-ray to EUV.}.

Such conditions are probably achievable within the framework of photoevaporation. Models for X-ray or FUV photoevaporation can achieve mass-loss rates $\gtrsim 10^{-8}\,M_{\sun}\,\mathrm{yr}^{-1}$ \citep[e.g.][]{Owen_2012,Nakatani_2018a,Nakatani_2018b,Picogna_2021}.
Moreover, although the simplest energetic arguments would suggest mass is lost from $r \gtrsim r_G $, hydrodynamic simulations demonstrate that significant mass loss extends inwards to $r = 0.1-0.2 \,r_G$ \citep[e.g.][]{Liffman_2003,Font_2004,Dullemond_2007,Alexander_2014}, exactly as we prefer here.
The FUV photoevaporation simulations of \citet{Komaki_2021}, while having a local peak around the typical cut-off at $\sim0.2-0.3\,r_G$, even suggest mass loss in as far as $0.05\,\mathrm{au}$.
Finally, \citet{Sellek_2022} argued that assuming the cooling is dominated by atomic emission lines, at typical luminosities, the X-rays which could most effectively drive a wind were $\sim 500\,\mathrm{eV}$ (though this value may shift to higher energies at higher luminosities) and these should achieve $\tau\sim1$. Winds resulting from X-ray photoevaporation should hence indeed be optically thick to the $\sim 250\,\mathrm{eV}$ photons that ionize Ar, but potentially not to the $\sim 870\,\mathrm{eV}$ photons that ionize Ne.

However, attributing the wind to photoevaporation is not without caveats.
Firstly, such high mass-loss rates as we infer are not universal to photoevaporation simulations with other treatments finding lower values \citep{Komaki_2021,Sellek_2024b}, albeit at X-ray luminosities somewhat below that used in our best-fitting hiLX\_TCha\_sUV spectrum (which might be expected to drive a higher mass-loss rate). This is in agreement with studies of disk population synthesis and evolution on secular timescales, where many authors have argued that the higher photoevaporation rates cannot be sustained across the whole population \citep{Somigliana_2020,Sellek_2020b,Emsenhuber_2023,Appelgren_2023} in order to explain the lifetimes of disks and the observation of disks with large amounts of (dust) mass depletion. The most statistical constraint was calculated by \citet{Alexander_2023} who showed that the observed distribution of accretion rates is incompatible with a simulated population if the photoevaporation rate exceeds $\sim10^{-9}\,M_{\sun}\,\mathrm{yr^{-1}}$. 
This arises since once the photoevaporation rate exceeds the accretion rate, photoevaporation is expected to quickly open a gas cavity around $r_G$ \citep{Clarke_2001} leading to the rapid dispersal of the inner disk and the quenching of accretion. 

Our inferred mass-loss rate $\gtrsim 10^{-8}\,M_{\sun}\,\mathrm{yr}^{-1}$ for T Cha would be somewhat higher than its measured accretion rate $\sim 4 \times 10^{-9}\,M_{\sun}\,\mathrm{yr}^{-1}$ and is thus subject to the above issue.
While in principle this could be rescued if the photoevaporative wind is less extended than assumed here (which would lower our estimate of the mass-loss rate since it diverges to large radii), the required extent would be so small as to make it unlikely that we would detect an extension in the [\ion{Ne}{2}] emission. 
Therefore, it may be that we are observing the disk during the final stages of dispersal outlined above. Such a scenario would be supported by the changes in the SED which suggest a significant reduction in the inner disk mass \citep{Xie_2023}, although there is no clear evidence for a gas cavity in the disk around T Cha \citep{Wolfer_2023}.

Finally, although in photoevaporation models some mass-loss continues as far as $0.1\,r_G$, it is not clear that the power-law profile of density assumed in the self-similar model also extends thus far: if the density gradient becomes shallower than $r^{-1}$ it will become hard to match the constraints on column density we infer from the observations.
The fact that the Ne line ratios predicted by \citet{Ercolano_2010} are closest to those produced by our $r_{\rm in}=r_G$ models may suggest that our models with a smaller inner radius do somewhat overestimate how close to the star EUV is absorbed compared to photoevaporation simulations.

MHD winds may solve many of the above problems for the wider protoplanetary disk population: as a transition disk with only weakly blueshifted lines, T Cha is not representative of this overall population. The blueshift of the [\ion{O}{1}] line is found to reduce with the accretion luminosity of the source \citep{Banzatti_2019}; therefore many disks with higher accretion rates have lines that imply wind velocities higher than can be achieved in photoevaporative winds. Moreover, the broad components of the lines in these disks can only be obtained by Keplerian broadening at $\lesssim 0.5\,\mathrm{au}$ \citep{Simon_2016,Banzatti_2019}, too close to the star to be produced in a photoevaporative wind. Likewise, higher accretion rate sources frequently show [\ion{Ne}{2}] HVCs representative of a jet \citep{Pascucci_2020}. These line features are thus attributed to inner MHD winds/jets. MHD winds may thus be one way to explain the breadth of the [\ion{Ne}{2}] line, which we were not able to reproduce adequately in this work.

MHD winds also provide benefits as they can extract angular momentum from disks and drive accretion: they thus form an attractive alternative to viscous accretion \citep{Manara_2023} - especially in the context of several developing lines of evidence for weak turbulence at large radii in protoplanetary disks \citep{Rosotti_2023}. Photoevaporation on the other hand does not produce accretion torques so it is unclear if accretion can be explained within a purely photoevaporative context.
Moreover, extended MHD winds can sustain total mass-loss rates somewhat higher than the accretion rate onto the star \citep[e.g.][]{Ferreira_1995}, potentially helping to explain the inferred wind mass-loss rate being higher than the measured accretion rate \citep{Cahill_2019}.
They may also sustain rapid gas accretion through a cavity \citep[e.g.][]{Martel_2022}, allowing accretion to continue during disk dispersal.
Global MHD simulations \citep[e.g.][]{Rodenkirch_2020} generally suggest that magnetic fields boost the mass-loss rate over those obtained with photoevaporation alone, which may further explain the high mass-loss rates we find for T Cha.

Overall, while MHD winds are attractive from the population perspective, we can more or less adequately explain the observations of T Cha with a photoevaporation model and it is not clear that MHD winds are needed to explain this disk (a transition disk with a large dust cavity potentially near the end of its lifetime). This is consistent with one suggested picture of MHD winds dominating for most of the disk lifetime before giving way to photoevaporation towards the end for the final disk dispersal \citep{Pascucci_2020,Kunitomo_2020,Weder_2023}.
Nevertheless, it would be useful for future work to more explicitly model MHD winds in a similar framework to that we use here in order to establish how the observational signatures such as the relative extents of the [\ion{Ne}{2}] and [\ion{Ar}{2}] emission would differ from photoevaporation.

\subsubsection{What absorbs the soft X-rays and EUV?}
We reiterate that $r_{\rm in}$ is simply a radius at which we truncate our density grid and that realistically some material, for example the inner disk's hydrostatic atmosphere, must lie between the modeled wind and the star.
Other contributions could come from a separate inner wind component (driven by magnetic forces) or accretion streams.
In any case, so long as the column density contributed by this material is not significant compared to that in our wind model, our results would not be strongly affected.

Absorption from material close to the star can also be traced at shorter wavelengths \citep{Edwards_2006,LopezMartinez_2015,Erkal_2022}. In particular, the Lyman $\alpha$ profile has been used to distinguish between absorption in an outflow (P-Cygni profiles), or accretion streams (inverse P-Cygni profiles) \citep{Arulanantham_2021}. T Cha shows a P-Cygni--like profile to which \citet{Arulanantham_2023} fit a model for the absorption with a velocity of $-101\,\mathrm{km\,s^{-1}}$. This suggests that there is indeed a fast outflow producing some degree of absorption close to the star that could be responsible for screening the outer wind (rather than accretion streams or a static atmosphere). Achieving such velocities certainly requires the outflow to be magnetically launched and suggests a Keplerian radius $\sim 0.01\,r_G$. An outflow is also seen in absorption in the \ion{C}{2} $1335$ \r{A} line \citep{Xu_2021}, albeit at smaller velocities of $-15\,\mathrm{km\,s^{-1}}$.

While none of T Cha's emission lines display a HVC \citep{Pascucci_2020}, it is part of the transition disk sample for which \citet{Rota_2024} suggest free-free emission may originate in a jet. Jets are typically extended away from the disk and, depending on where exactly they launch may or may not be able to screen the outer disk. However, no source displays both a HVC and LVC in [\ion{Ne}{2}] \citep{Pascucci_2020} which may support the possibility that mass-loss associated with a jet can screen the outer disk in some sources (those with a higher accretion rate). In this case, T Cha's jet, if present, must have weakened sufficiently that it no longer screens the outer disk, thus allowing the $\sim 870\,\mathrm{eV}$ photons that ionize Ne can now penetrate to the outer disk.
If the free-free emission originates in a jet, \citet{Rota_2024} estimate the mass-loss rate at $\sim5\times10^{-10}\,M_{\sun}\,\mathrm{yr^{-1}}$.
Assuming an inner radius equal to the magnetospheric truncation radius at a few stellar radii \citep[$R_{\rm T\,Cha}=1.3\,R_{\sun}$,][]{Olofsson_2011}, i.e. $r_{\rm in}\sim 0.003\, r_G$, Equation \ref{eq:N_values} would imply a column density on the order of $3\times10^{19}\,\mathrm{cm^{-2}}$. This is an order of magnitude below the column density required to screen the $250\,\mathrm{eV}$ photons and suggests that a jet cannot be responsible for absorbing the soft X-rays.

Evidence for an inner-wind-shielding scenario affecting the mid-IR lines for SZ Cha was recently presented \citet{Espaillat_2023} who found a strong decrease in the ratio of the [\ion{Ne}{3}] and [\ion{Ne}{2}] lines (accompanying a decrease in both lines' fluxes) between Spitzer data and MIRI-MRS data. By comparison to variability in SZ Cha's ${\rm H \alpha}$ spectrum, they suggest that an inner wind was suppressed at the time of the Spitzer measurements - allowing EUV to ionize the outer wind - while in the latest epoch the wind has returned, thus screening out EUV and leading to only X-ray ionization in the outer wind. Such a picture would be consistent with, for example, the left-hand panel of Figure \ref{fig:NeRatio_compare} where for several spectra we see the ratio cross from brighter [\ion{Ne}{3}] to brighter [\ion{Ne}{2}] as the inner radius is reduced. A similar inner wind could thus conceivably contribute to absorbing the soft X-ray and EUV photons for T Cha.

\citet{Pascucci_2020} presented an evolutionary sketch where full disks, with higher accretion rates, would host a dense inner wind which screens even the $\sim 1\,\mathrm{keV}$ X-rays from the outer disk, preventing the detection of an LVC, while disks with a cavity may host only a tenuous inner wind. Our interpretation of the data for T Cha suggests that while it likely hosts some sort of inner wind, it can at most absorb only the soft X-rays, with the inner parts of a dense photoevaporative wind themselves providing most of the screening.

\subsection{Further prospects for resolving wind emission}
Given the evolutionary and demographic implications of the high mass-loss rates inferred in our work, it is important to understand if similar conclusions would be drawn across the protoplanetary disk population, or if T Cha is something of an outlier, perhaps due to its higher stellar mass ($1.5\,M_{\sun}$) compared to most T Tauri stars, or its status as a more evolved disk.
Our GO 2260 program has also observed V4046 Sgr, and programs GO 1676 \citep{Espaillat_2021} and GO 3983 \citep{Thanathibodee_2023} will constrain [\ion{Ne}{3}] to [\ion{Ne}{2}] ratios (particularly in lower accretors) - a good measure of whether the mass-loss rate is low enough for EUV to penetrate and ionize the wind. Such efforts should help establish whether other more evolved disks appear to have dense photoevaporative winds.
In addition, since photoevaporation is known to be insufficient to explain the broader or more strongly blueshifted line profiles observed for many disks, and in order to test the evolutionary perspective outlined by \citet{Pascucci_2020}, it would also be informative to compare to a sample of younger, full disks to see if the [\ion{Ne}{2}] is more compact as would be expected if the hard X-rays are screened out by a more substantial inner wind.

A recent parallel advance is the use of the Multi Unit Spectroscopic Explorer (MUSE) instrument on the Very Large Telescope (VLT) to perform high-spatial-resolution spectral mapping of [\ion{O}{1}] emission \citep{Fang_2023b,Flores-Rivera_2023}.
For example, for another transition disk which has a low-velocity [\ion{Ne}{2}] component - TW Hya - these efforts yielded the first conclusive detection of a blueshift ($-0.8\,\mathrm{km\,s^{-1}}$) in the [\ion{O}{1}] \citep{Fang_2023b}. Since 80 per cent of the emission originates within 1 au of the star, these observations were considered consistent with the magnetohydrodynamic models of \citet{Wang_2019}.
However they can also be fit with a photoevaporative wind \citep{Rab_2023} albeit with the emitting region of the [\ion{O}{1}] line lying mostly in the bound, static, atmosphere just inside of the wind \citep[thus explaining its lack of a significant blueshift without the need for a dust cavity, the explanation proposed by][]{Pascucci_2011}. [\ion{O}{1}] emission was also described as predominantly originating in an almost hydrostatic atmosphere in the photoevaporative models by \citet{Nemer_2023}.
This restricted emitting region is due to higher excitation temperature of the [\ion{O}{1}] line, which can only be produced in the hottest, innermost regions where the EUV penetrates \citep{Ercolano_2016}.
This would suggest a picture where a photoevaporative wind extends in to $\sim 1\,\mathrm{au}$ much like our models suggest fits T Cha.
Comparing the two complementary approaches to resolving line emission would be a fruitful avenue for future work, and MIRI-MRS observations of TW Hya are included within the GTO program MINDS \citep[1282][]{Henning_2017}.

\section{Conclusions}
\label{sec:conclusions}
\citet{Bajaj_2024} recently presented the first evidence of spatially extended emission from [\ion{Ne}{2}] in a low-velocity disk wind from T Cha, as well as the first detection of [\ion{Ar}{3}] emission, making it the first case where [\ion{Ne}{2}], [\ion{Ne}{3}], [\ion{Ar}{2}] and [\ion{Ar}{3}] are all confidently detected together. In this work we have modeled these emission lines using a simple prescription for a thermal disk wind \citep{Sellek_2021} post-processed with a photoionization code \citep[\textsc{mocassin}][]{Ercolano_2003,Ercolano_2005,Ercolano_2008} in order to establish the emitting regions and luminosities of each line. We then used \textsc{MIRISim} \citep{Klaassen_2021} - with some modifications - to create synthetic images for comparison to the data.
The main conclusions from this effort are that
\begin{itemize}
    \item Based on the III-II line ratio of both the Ne and Ar emission, the degree of ionization of the material must be low, ruling out the possibility that ionization is dominated by EUV. This in turn excludes the prospect of a low--mass-loss rate, low-density wind for T Cha, as might be expected in models of EUV photoevaporation (Figure \ref{fig:IIItoII}).
    \item Based on the [\ion{Ne}{2}] and [\ion{Ne}{3}] line fluxes, T Cha instead hosts a dense wind irradiated by a luminous X-ray spectrum (Figure \ref{fig:NeRatio_compare}). This suggests a mass-loss rate on the order of $10^{-8}\,M_{\sun}\,\mathrm{yr^{-1}}$.
    \item The synthetic imaging suggests that were the [\ion{Ne}{2}] emission coming exclusively from scales greater than the $\sim 30\,\mathrm{au}$ mm cavity, it would be easily resolvable and its extent would trace the inner radius of the wind which is heated by the EUV. Conversely, for inner radii less than the gravitational radius (which for T Cha is approximately 13 au), the extent traces extended, X-ray--ionized gas and so becomes only weakly dependent on the inner radius (Figure \ref{fig:HWHM_NeII}). Therefore, from the $\sim 10\,\mathrm{au}$ spatial extent of the [\ion{Ne}{2}] emission alone, it is hard to place tight constraints on the inner radius, though this requires that a wind is launched - and ionized by hard X-rays - on scales of at least $12\,\mathrm{au}$. This constrains the column density inside $12\,\mathrm{au}$ to be $<10^{22}\,\mathrm{cm^{-2}}$. Moreover, our models - which reproduce the observed extent well - have significant emission out to nearly $100\,\mathrm{au}$ thus suggesting a significantly extended wind.
    \item In order to reproduce the slightly fainter [\ion{Ar}{2}] than [\ion{Ne}{2}] emission (Figure \ref{fig:ArRatio_compare}), along with the less-extended [\ion{Ar}{2}] emitting region inferred from its lack of extension over the PSF (Figures \ref{fig:HWHM_ArII} \& \ref{fig:HWHM_vs}), we require that on scales $\gtrsim 12\,\mathrm{au}$ only hard--X-ray photons $>870\,\mathrm{eV}$ are able to penetrate and ionize the outer parts of the wind, while soft--X-ray photons $\sim 250\,\mathrm{eV}$ (and EUV) are screened by the inner parts of the wind. This is equivalent to requiring a column density of $\gtrsim 3\times10^{20}\,\mathrm{cm^{-2}}$ on $12\,\mathrm{au}$ scales. For reasonable mass-loss rates, this in turn requires wind material down to quite small inner radii $\lesssim 0.1 \,r_G \approx 1.3\,\mathrm{au}$.
    \item Moreover, the observed position angle of the [\ion{Ne}{2}] emission matches well that predicted in the synthetic images, which is close to perpendicular to the disk major axis.
    \item High spectral resolution line profiles of [\ion{Ne}{2}] (Figures \ref{fig:profile_NeII} \& \ref{fig:profile_NeII_slit}) - and also [\ion{O}{1}] (Figure \ref{fig:profile_OI}) - also support the wind having a small inner radius in order to reproduce their breadth.
\end{itemize}

In particular we highlight that the combination of [\ion{Ar}{2}] and [\ion{Ne}{2}] provides the best constraints on the column density absorbing photoionizing X-rays and thus on the mass-loss rate and innermost extent of the wind.

The solution we find of a mass-loss rate of $\sim3\times10^{-8}\,M_{\sun}\,\mathrm{yr^{-1}}$ and an inner radius of $r_{\rm in}\approx0.1 \,r_G \approx 1.3\,\mathrm{au}$ are suggestive of photoevaporation if we assume the higher mass-loss rates of some X-ray--driven \citep{Owen_2012,Picogna_2019,Ercolano_2021,Picogna_2021} or FUV-driven \citep{Nakatani_2018a,Nakatani_2018b} models.

On the other hand, high mass-loss rates, particularly those in excess of the accretion rate, could suggest that magnetic fields assist in wind launching.
Moreover, the broad [\ion{Ne}{2}] line profile and presence of absorption at high velocities in the Lyman $\alpha$ \citep{Arulanantham_2023} could suggest of an inner magnetocentrifugal wind. For such an inner disk wind not to increase the total wind column density significantly and lead to compact [\ion{Ne}{2}], its mass-loss rate would have to be somewhat lower and it would likely struggle to simultaneously reproduce the observed line fluxes. Thus we still require a dense extended outer disk wind, suggesting a role for thermal pressure gradients (i.e. photoevaporation) assisting in the launch at large radii either way.

\section*{Acknowledgments}
We thank the reviewers for their constructive reports and suggestions that improved the clarity of our results. ADS also thanks Christian Rab for useful discussions.

ADS thanks the Science and Technology Facilities Council (STFC) for a Ph.D. studentship for project 2277492 as part of Training Grant ST/T505985/1 and was also supported by funding from the European Research Council (ERC) under the European Union’s Horizon 2020 research and innovation programme (grant agreement No. 1010197S1 MOLDISK).
This work has also been supported by the European Union's Horizon 2020 research and innovation programme under the Marie Sklodowska-Curie grant agreement No 823823 (DUSTBUSTERS) and ADS thanks all those at the University of Arizona for their hospitality during the associated secondment at the commencement of this work.
NSB and IP were funded by the NASA/STScI GO grant JWST-GO-02260.001.
CJC acknowledges support from the Science \& Technology Facilities Council (STFC) Consolidated Grant ST/W000997/1 \& RA acknowledges funding from the Science \& Technology Facilities Council (STFC) through Consolidated Grant ST/W000857/1.
GB has received funding from the European Research Council (ERC) under the European Union’s Horizon 2020 Framework Programme (grant agreement no. 853022, PEVAP). 

This work was performed using resources provided by the Cambridge Service for Data Driven Discovery (CSD3) operated by the University of Cambridge Research Computing Service (www.csd3.cam.ac.uk), provided by Dell EMC and Intel using Tier-2 funding from the Engineering and Physical Sciences Research Council (capital grant EP/T022159/1), and DiRAC funding from the Science and Technology Facilities Council (www.dirac.ac.uk).

This work is also based on observations made with the NASA/ ESA/CSA James Webb Space Telescope. The data were obtained from the Mikulski Archive for Space Telescopes at the Space Telescope Science Institute, which is operated by the Association of Universities for Research in Astronomy, Inc., under NASA contract NAS 5-03127 for JWST. The observations are associated with JWST GO Cycle 1 program ID 2260. The JWST data used in this paper can be found in MAST: \href{http://dx.doi.org/10.17909/dhmh-fx64}{10.17909/dhmh-fx64}.
The synthetic detector images corresponding to the favoured $r_{\rm in}=0.1\,r_G$ models (with and without a cavity) plus those of the simulated standard star are available via a Zenodo repository at \href{http://doi.org/10.5281/zenodo.10409615}{10.5281/zenodo.10409615}. Other models can be shared upon request to the corresponding author.

This work also benefited from the Core2disk-III residential program of Institut Pascal at Universit\'e Paris-Saclay, with the support of the program ``Investissements d’avenir'' ANR-11-IDEX-0003-01.

\begin{software}
    NumPy \citep{vanderwalt_2011}, 
    SciPy \citep{Virtanen_2020},
    Matplotlib \citep{Hunter_2007},
    MOCASSIN \citep{Ercolano_2003,Ercolano_2005,Ercolano_2008},
    MIRISim \citep{Klaassen_2021},
    JWST Pipeline \citep{JWST_pipeline},
    CASA \citep{McMullin_2007}
\end{software}


\bibliography{biblio}{}
\bibliographystyle{aasjournal}


\appendix
\section{Modifications to MIRISim}
\label{appendix:MIRISim}
As the latest version of \textsc{MIRISim} was released prior to the launch of JWST, the Calibration Data Products (CDP) it uses are from the ground-based testing, and for several effects have been superseded by inflight measurements.
For a better direct comparison with the data we therefore downloaded the new CDP files for the following calibrations: DISTORTION (9B.05.07), FLAT (08.01.00), FRINGE (Channels 3\&4 A/B/C only: 8P.02.00/9B.00.00/8P.01.06), PHOTOM (9B.04.08) and PSF (09.02.00) and modified \textsc{MIRISim} to read these files instead of its defaults.

We also modified \textsc{MIRISim} to set the observation time keywords in the output file headers to the date of the real observations (as otherwise the pipeline would default to using the ground-based calibration files when analyzing our synthetic images). Moreover, as of version 1.11 of the pipeline, the ``EXPMID'' keyword is required to calculate the time-dependent photometric calibration and we modified \textsc{MIRISim} to include this in the headers.

In testing our method we found that when \textsc{MIRISim} interpolates an input FITS file (as used for the line emission) onto its internal grid, flux on scales smaller than a pixel - as expected for an unresolved wind - was not well-conserved. We therefore implemented in \textsc{MIRISim} our own 2-dimensional histogram interpolation method - where the luminosity in each cell in the input grid is placed into the nearest neighbour cell in the \textsc{MIRISim} grid (or shared equally between cells in the case of a tie) - which we found corrected this issue adequately.


We also discovered that \textsc{MIRISim} did not apply the PSF convolution to extended sources when simulating MIRI-MRS. We thus include in our interpolation a step to upsample the PSF over the field of view at the resolution of the input FITS file, and apply the convolution to the data before the histogram rebinning is done. This is done in each wavelength channel of the input file. Finally, the contribution of the convolved, rebinned emission from each channel is added to all wavelength channels of MIRI-MRS according to the line spread function. An equivalent PSF upsampling and convolution method was applied to the disk model.

\section{Cross sections of the wind for different parameters}
\label{appendix:additionalmaps}
\begin{figure*}[p]
    \centering
    \includegraphics[angle=90,origin=c,height=\linewidth]{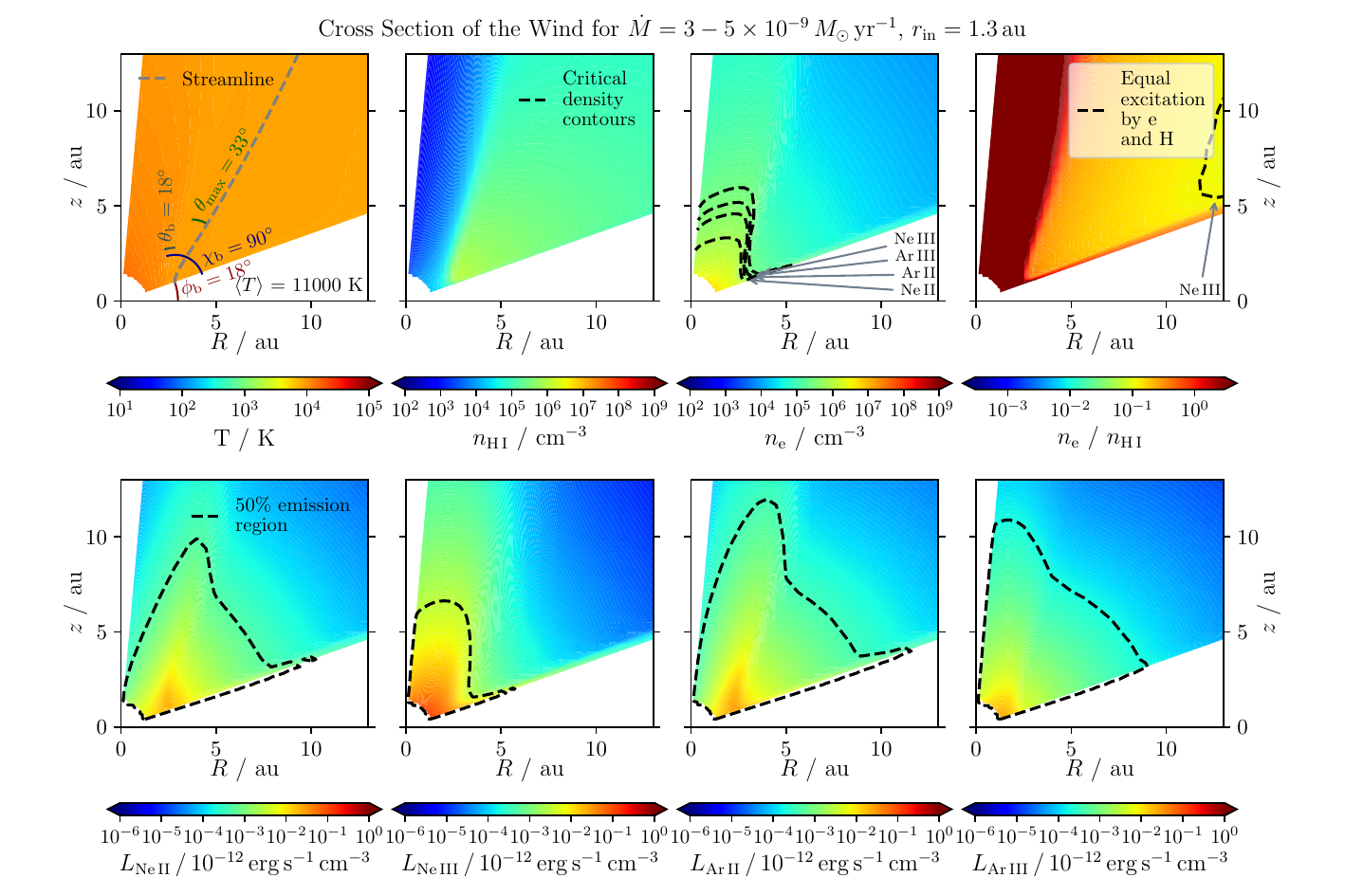}
    \caption{As with Figure \ref{fig:Rz_maps} but for a gas density 10 times lower.}
    \label{fig:Rz_maps_Mdot8}
\end{figure*}

Figure \ref{fig:Rz_maps_Mdot8} shows that for a lower mass-loss rate, the wind is more highly ionized. Collisions with electrons thus dominate over neutral H throughout the emitting region for all lines of interest. The [\ion{Ar}{2}] emission is comparably extended to that of the [\ion{Ne}{2}]

\begin{figure*}[p]
    \centering
    \includegraphics[angle=90,origin=c,height=\linewidth]{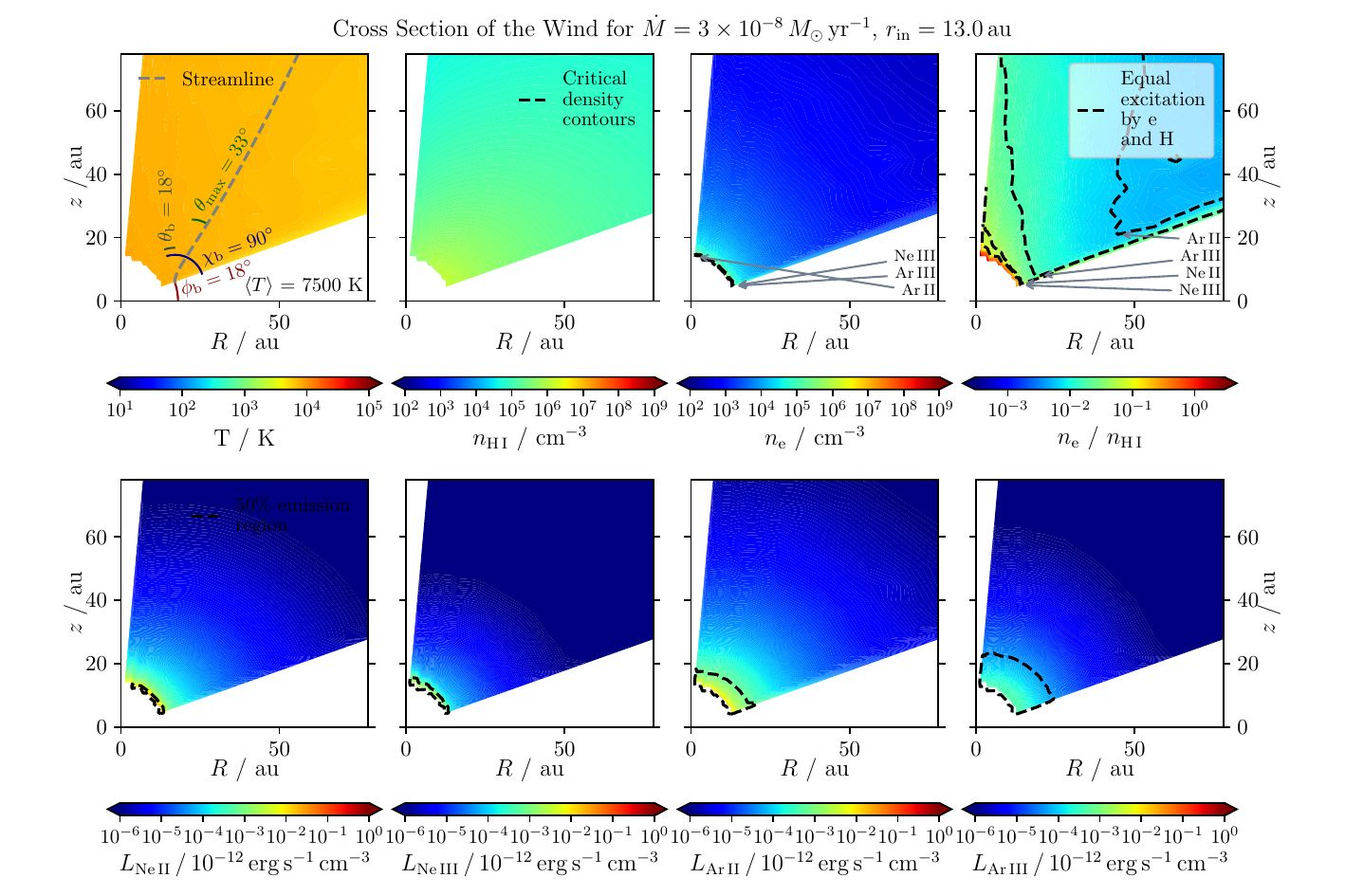}
    \caption{As with Figure \ref{fig:Rz_maps} but for $r_{\rm in}=r_G$. Note the change in scale of the panels to accommodate the larger scales on which the wind is ionized.}
    \label{fig:Rz_maps_in0}
\end{figure*}

Figure \ref{fig:Rz_maps_in0} shows that for a larger $r_{\rm in}$, the emitting regions are generally very thin and close to the inner radial boundary of the wind where the wind is ionized. This means that electrons dominate over neutral H throughout the emitting region for all lines of interest.
The [\ion{Ar}{2}] emission is now slightly more extended than that of the [\ion{Ne}{2}]

\section{Effect of collisional excitation by neutral H}
\label{appendix:collisions}
In Figures \ref{fig:Rz_maps}, \ref{fig:Rz_maps_Mdot8} and \ref{fig:Rz_maps_in0} we compare the critical densities (at the average gas temperature indicated in the first panel) for the excitation of the mid-IR noble gas lines by neutral H and electrons to the densities of these species determined from the photoionization radiative transfer.

For mass-loss rates $\lesssim 5\times10^{-9}\,M_{\odot}\,\mathrm{yr^{-1}}$ (Figure \ref{fig:Rz_maps_Mdot8}), we can see that the low gas densities and moderate ionization levels mean that neutral H never reaches the critical density, whereas the electrons do within the inner $3-5\,\mathrm{au}$ of the wind. This is a clear indication that in the inner parts of the wind, electrons are the more important colliders. The neutral H only starts to dominate once the X-ray ionization falls low enough, which happens outside $\sim 10\,\mathrm{au}$, while the majority of the line emission (assuming only electrons) comes from inside $10\,\mathrm{au}$; even then the first species for which neutral H comes a significant collider is \ion{Ne}{3} which presents the most compact emission under these conditions as it predominantly comes from the highly ionization EUV-heated region. Therefore, for all but the highest mass-loss rates we can be sure that the lack of neutral H as a collider cannot significantly affect our results as it only increases the excitation significantly in already very faint regions.

At the higher mass-loss rates that we prefer (Figure \ref{fig:Rz_maps}), however, the denser wind has somewhat lower ionization levels, since recombination is more effective and the ionizing radiation is more easily attenuated. Now, the collisional excitation with neutral H can dominate over that with electrons on similar scales to where the wind emission is mostly coming from. The Ar lines are the least affected since the emitting extent is limited more by the ionization: for [\ion{Ar}{2}] most emission originates within $\sim3\mathrm{au}$ while neutral H becomes important beyond $\sim6\,\mathrm{au}$.
Consequently, the Ar II/III luminosities should be robust predictions, and the Ar II emission should remain compact.
On the other hand, since the critical densities of the Ne lines due to neutral H are much closer to those with electrons than for the Ar lines, neutral H dominates most of the emitting region of the wind, save for the very inner edge where the photoionization by EUV is still significant.

In these cases, the line emission may be boosted by a factor up to $\frac{n_{\rm H}}{n_{\rm e}} / \frac{n_{\rm crit, H}}{n_{\rm crit, e}}$. However this is an upper limit: for example in the case of the [\ion{Ne}{3}] line the collisions with neutral H bring most of the emitting region into LTE, when the excitation (and hence line emission) saturates and is no longer sensitive to the collider densities. Moreover, the broader inclusion of such additional pathways for collisionally-excited emission would lead to additional cooling \citep{Sellek_2024b}, thus lowering the temperature of the outer regions of the wind somewhat. This would both reduce the line excitation and raise the critical densities with neutral H of the lines of interest, thus reducing the additional emission with respect to what might be estimated from these figures.

A similar picture is seen for a larger $r_{\rm in}$ (Figure \ref{fig:Rz_maps_in0}), but here the emission (assuming only electrons) was generally limited to the EUV-ionized region. This means that the X-ray-ionized regions contribute relatively less to the overall line flux than at smaller $r_{\rm in}$ and hence boosting the excitation in this region would have a more limited effect.

It is overall therefore likely that in denser environments neutral H can be an important collider and now that rates are available \citep{Yan_2024}, this should be considered more explicitly and self-consistently in future works in order to disentangle some of the opposing effects described above, at least for the densest winds.
This may in turn affect the accuracy with which our preferred model of a dense wind - with a high mass-loss rate and small inner radius - agrees with the data.
However, since a lot of our conclusions are based on ruling out the rest of the parameter space where neutral H is less significant on the basis of line ratios, we can satisfactorily argue that they will still not describe the data if neutral H is considered. Moreover, some of its effects would only strength our conclusions: for example, since in the dense winds collisional excitation by neutral H affects \ion{Ne}{2} much more than \ion{Ar}{2}, it should increase the emitting extent of [\ion{Ne}{2}] with respect to [\ion{Ar}{2}] and hence the dense winds would remain the most consistent with extended [\ion{Ne}{2}] emission but compact [\ion{Ar}{2}].


\end{document}